\documentclass[american,english,journal, onecolumn, letterpaper, 12pt, final]{IEEEtran}
\usepackage[T1]{fontenc}
\usepackage[latin9]{inputenc}
\usepackage[letterpaper]{geometry}
\usepackage{babel}
\usepackage{refstyle}
\usepackage{url}
\usepackage{amsthm}
\usepackage{amsmath}
\usepackage{amssymb}
\usepackage{fixltx2e}
\usepackage{graphicx}
\usepackage{esint}
\usepackage[unicode=true,
 bookmarks=true,bookmarksnumbered=false,bookmarksopen=false,
 breaklinks=false,pdfborder={0 0 0},backref=false,colorlinks=false]
 {hyperref}

\makeatletter

\AtBeginDocument{\providecommand\secref[1]{\ref{sec:#1}}}
\AtBeginDocument{\providecommand\figref[1]{\ref{fig:#1}}}
\AtBeginDocument{\providecommand\tabref[1]{\ref{tab:#1}}}
\AtBeginDocument{\providecommand\eqref[1]{\ref{eq:#1}}}
\AtBeginDocument{\providecommand\thmref[1]{\ref{thm:#1}}}
\AtBeginDocument{\providecommand\apxref[1]{\ref{apx:#1}}}
\providecommand{\tabularnewline}{\\}
\newcommand{\lyxdot}{.}

\RS@ifundefined{subref}
  {\def\RSsubtxt{section~}\newref{sub}{name = \RSsubtxt}}
  {}
\RS@ifundefined{thmref}
  {\def\RSthmtxt{theorem~}\newref{thm}{name = \RSthmtxt}}
  {}
\RS@ifundefined{lemref}
  {\def\RSlemtxt{lemma~}\newref{lem}{name = \RSlemtxt}}
  {}

\theoremstyle{plain}
\newtheorem{thm}{\protect\theoremname}

\usepackage{etoolbox}
\usepackage{siunitx}
\usepackage{fp}
\usepackage{cite}

\usepackage{microtype}
\usepackage{nccmath}

\newref{apx}{name=Appendix~, names=Appendices~}
\newref{thm}{name=Theorem~, names=Theorems~}
\newref{eq}{name=Eq.~, names=Eqs.~, refcmd=\textup{(\ref{#1})}}

\newref{sec}{name=Section~, names=Sections~}
\newref{tab}{name=Table~, names=Tables~}
\newref{fig}{name=Fig.~, names=Figs.~}
\newref{thm}{name=Theorem~, names=Thms.~}

\def\fleqn{\@fleqntrue} 

\usepackage{colortbl}
\newcolumntype{G}{>{\columncolor[gray]{0.9}}c}

\usepackage[absolute]{textpos}

\@ifundefined{showcaptionsetup}{}{%
 \PassOptionsToPackage{caption=false}{subfig}}
\usepackage{subfig}
\makeatother

\addto\captionsamerican{\renewcommand{\theoremname}{Theorem}}
\addto\captionsenglish{\renewcommand{\theoremname}{Theorem}}
\providecommand{\theoremname}{Theorem}

\begin{document}

\title{An Analytical Model of Packet Collisions\\
in IEEE 802.15.4 Wireless Networks}

\author{Matthias Wilhelm, Vincent Lenders{*}, and Jens B.~Schmitt\\
Department of Computer Science, TU Kaiserslautern\\
67663 Kaiserslautern, Germany\\
{*}armasuisse Science and Technology, 3602 Thun, Switzerland}
\maketitle
\begin{abstract}
Numerous studies have shown that concurrent transmissions can help
to boost wireless network performance despite the possibility of packet
collisions. However, while these works provide empirical evidence
that concurrent transmissions may be received reliably, existing signal
capture models only partially explain the root causes of this phenomenon.
We present a comprehensive mathematical model for MSK-modulated signals
that makes the reasons explicit and thus provides fundamental insights
on the key parameters governing the successful reception of colliding
transmissions. A major contribution is the closed-form derivation
of the receiver bit decision variable for an arbitrary number of colliding
signals and constellations of power ratios, time offsets, and carrier
phase offsets. We systematically explore the factors for successful
packet delivery under concurrent transmissions across the whole parameter
space of the model. We confirm the capture threshold behavior observed
in previous studies but also reveal new insights relevant to the design
of optimal protocols: We identify capture zones depending not only
on the signal power ratio but also on time and phase offsets.
\end{abstract}
\begin{textblock}{0.8}[0.5,0.5](0.5,0.94)
\hrule\vskip 2pt
\centering\noindent\footnotesize

This is the accepted version of the article (with additional figures), to appear in IEEE Transactions on Wireless Communications.\\
For the final version visit \url{http://dx.doi.org/10.1109/TWC.2014.2349896} once it becomes available.
\vskip 4pt
\copyright\,2014 IEEE. Personal use of this material is permitted. Permission from IEEE must be obtained for all other uses, in any current or future media, including reprinting/republishing this material for advertising or promotional purposes, creating new collective works, for resale or redistribution to servers or lists, or reuse of any copyrighted component of this work in other works.
\end{textblock}

\section{Introduction}

Conventional wireless communication systems consider packet collisions
as problematic and try to avoid them by using techniques like carrier
sense, channel reservations (virtual carrier sense, RTS\slash{}CTS
handshakes), or arbitrated medium access (TDMA, polling). The intuition
is that concurrent transmissions cause irreparable bit errors at the
receiver and render packet transmissions undecodable. However, researchers
have found that this notion is too conservative. If the power of the
signal of interest exceeds the sum of interference from colliding
packets by a certain\emph{ }threshold, packets can in general still
be received successfully despite collisions at the receiver. This
effect, referred to as the \emph{capture effect} \cite{CaptureFM},
has been explored extensively and validated in many independent practical
studies on various communication systems such as IEEE 802.11 \cite{08:MultiCapture:JSAC,UnderstandRFInterf,04:WLANCapture:ICNP,Capture11a}
and IEEE 802.15.4 \cite{10:154CaptureModel:ISWPC,InfModel,ExpConcur}. 

Over the past years, the view on packet collisions has therefore changed
considerably. Since it is possible for some or even all packets in
a collision to survive, there are opportunities to increase the overall
channel utilization and to improve the network throughput by designing
protocols that carefully select terminals for transmitting at the
same time \cite{CMAC,08:HarnessExposed:NSDI}. The benefits and potential
performance improvements of concurrent transmission are not just of
theoretical interest but have been demonstrated practically and adopted
in application areas such as any-cast \cite{10:AMAC:SenSys,ACKCollisions},
neighbor counting \cite{14:CountCC:IPSN}, or rapid network flooding
\cite{13:Splash:NSDI,Glossy,FlashFlood,13:ScaleFlood:ToN,Triggercast},
especially in the context of wireless sensor networks (WSNs).

Although protocols that exploit concurrent transmissions have shown
the potential to boost the overall performance of existing wireless
communication systems, their success cannot be explained with capture
threshold models based on the Signal to Interference and Noise Ratio
(SINR) alone. Recent studies have shown that, while the relative signal
powers of colliding packets indeed play an important role in the reception
probability, other factors are also of major importance. For example,
several experimental studies report that the relative timing between
colliding packets has a significant influence on the reception probability
\cite{Capture11a,CaptureAdHoc}. Others report that the coding \cite{DelayCap}
or packet content \cite{10:AMAC:SenSys} may also greatly influence
the reception performance in the presence of collisions. Further factors
such as the carrier phase offset between a packet of interest and
colliding packets also need to be considered \cite{11:Wireless-Manip:ESO}.

In this paper, we strive to provide a comprehensive model accounting
for all these factors, focusing on packet collisions in IEEE 802.15.4
based WSNs. Such a model will allow protocol designers to better understand
the root causes of packet reception and exact conditions under which
concurrent transmissions actually work, and thus to design optimal
protocols based on these factors. While previous studies \cite{UnderstandRFInterf,InfModel,05:Capture:EmNetS,TalkTogether,ModelGlossy}
also looked at factors that determine the success of concurrent packet
reception, these works are either based on practical experiments and
have therefore led to empirical models that cannot be generalized
easily, or derived simplified models that do not account for all impact
factors. This work advances the field by providing a unified analytical
model accounting for the major factors identified above (see also
\secref{Params}). Our model ($\rightarrow$~\secref{Model}) is
based on a mathematical representation of the physical layer using
continuous-time expressions of the IQ signals entering the receiver's
radio interface. This fundamental and comprehensive model allows to
represent an arbitrary number of colliding packets as a linear superposition
of the incoming signals. 

A major contribution of this work is a closed-form analytical representation
of the bit decision variable at an optimal receiver's demodulator
output based on these IQ signals ($\rightarrow$~\secref{Analysis}).
This result enables the deterministic computation of the bit demodulation
decision and hence to compute the actual performance of concurrent
transmissions for any colliding parameter constellations. Having a
bit-level model of reception is not only beneficial for the comprehension
of the collision process, it also contributes to application areas
where a precise bit-level analysis is needed, such as partial packet
reception \cite{PPR}, understanding bit error patterns in low-power
wireless networks \cite{BitErrDistLPWN,12:RedCliff:INFC}, or signal
manipulation attacks at the physical layer \cite{11:Wireless-Manip:ESO}.

Using our model, we explore the parameter space of the reception of
MSK-modulated colliding packets considering both uncoded and Direct
Sequence Spread Spectrum (DSSS) based systems ($\rightarrow$~\secref{Evaluation}),
analyzing the influence of the parameters on the resulting packet
reception ratio (PRR) for concurrent transmissions. While the analysis
shows that our model agrees with experimental results in the literature,
it also provides much more detailed insights into the performance
characteristics of protocols that exploit collisions \cite{13:Splash:NSDI,10:AMAC:SenSys,Glossy,FlashFlood,13:ScaleFlood:ToN,Triggercast}.
In particular, we show that the good performance of these protocols
should be attributed equally to coding (e.g., DSSS) and power capture.
In addition, based on our analysis we identified parameter constellations
where concurrent transmissions work reliably. We therefore propose
a generalization of the traditional capture threshold model based
on the power ratios towards a \emph{capture zone. }Capture zones result
from the model insight that reception success does not depend on the
power ratio between interfering signals alone, but on the time and
phase offsets of sender and receiver as well.

To show the validity and accuracy of our model, we implemented and
experimented with an application that is strongly dependent on physical
layer characteristics, the reception of unsynchronized signals. We
performed this experiment with two widely used commercial IEEE 802.15.4
receiver implementations (TI CC2420 and Atmel AT86RF230) to demonstrate
that our results are receiver-independent ($\rightarrow$~\secref{Experiments}).
The results validate our claim that our model accurately captures
the behavior of realistic receivers in the face of concurrent transmissions.
Finally, we discuss parameter settings for an optimal protocol design
($\rightarrow$~\secref{Discussion}).

\section{Impact Factors}

\label{sec:Params}Different factors influence the probability of
a successful reception under collisions. This section discusses the
main factors that have been identified in the literature. Subsequently,
we consider them jointly in our mathematical model to predict the
outcome of concurrent transmissions.

\subsubsection*{Power ratio}

The signal power is a crucial factor for successful reception in general,
and it plays a major role in the reception under collisions as well.
SINR-based models are widely used to model the packet reception in
a shared medium, for example in the Physical Model \cite{GuptaKumar}
and its variants \cite{ModelInterf,IntfModels}. The classical SINR
model states that a stronger signal is received if its signal power
$P_{s}$ exceeds the channel noise $P_{n}$ and the sum of interfering
signal powers $\sum_{i}P_{i}$ by a given threshold, i.e., 
\[
\frac{P_{s}}{P_{n}+\sum_{i}P_{i}}>\delta_{\mathrm{SINR}}.
\]
This simple model is accurate for uncorrelated interfering signals
such as additive white Gaussian noise (AWGN). However, when the interference
is correlated (such as colliding packets), this model is not always
accurate and further factors must be considered \cite{UnderstandRFInterf,Capture11a,CaptureAdHoc}.

\paragraph*{Signal timing}

The relative timing of colliding packets greatly influences the reception
process. This is because the receiver locks onto a packet during the
synchronization phase at the start of the transmission. If a stronger
signal arrives later, it captures the receiver and disturbs the first
packet reception, and both packets in the collision are lost. Thus,
in packet radios, power capture alone is not sufficient for successful
reception, rather the receiver must be synchronized and locked onto
the captured signal as well. Several research contributions analyze
possible collision constellations and their effect on packet reception
\cite{Capture11a,CaptureAdHoc}, and propose a new receiver design
that releases the lock when a stronger packet arrives, discards the
first and receives the second packet, the so-called \emph{message-in-message
(MIM) capture} \cite{Capture11a,05:Capture:EmNetS}. Subsequent works
apply these insights to improve network throughput. For example, Manweiler
et al.~\cite{Reorder} propose collision scheduling to ensure that
MIM is leveraged, thus increasing spatial reuse.

\paragraph*{Channel coding}

A further factor that influences packet reception success is bit-level
coding. For example, in DSSS systems a group of $b$ bits is encoded
into a longer sequence of $B$ chips \cite{07:DigiCom:book}. The
benefit of this approach is that resilience to interference is increased
because the chipping sequences can be cross-correlated at the receiver,
which effectively filters out uncoded noise. However, DSSS systems
require interfering signals to be uncorrelated, e.g., signals without
coding or with orthogonal chipping sequences (as in CDMA), to achieve
their theoretical coding gain. Another possibility is a sufficient
time offset between interfering packets with the same coding; this
phenomenon is known as \emph{delay capture} \cite{DelayCap}. As networking
standards such as IEEE 802.11 and IEEE 802.15.4 generally use DSSS
with identical codes for all participants, existing experimental works
on collisions and capture observe the effects of DSSS implicitly.

\paragraph*{Packet contents}

Experimental results show that packets with identical payload and
aligned starting times result in good reception performance and reduced
latency in broadcast scenarios. For example, Dutta et al.~\cite{10:AMAC:SenSys}
show that short packets can be received in such collisions with a
PRR over 90\,\%, thus enabling the design of an efficient receiver-initiated
link layer. Similarly, the latency of flooding protocols widely used
in WSNs can be greatly reduced \cite{Glossy,13:ScaleFlood:ToN}. In
these works, experiments in IEEE 802.15.4 networks reveal that the
tolerable time offset between concurrent messages is small (approx.~500\,ns),
which adds challenges to protocol design and implementation. These
insights also show that capture and packet synchronization alone are
not sufficient to explain the performance of these protocols, and
bit-level modeling that also includes signal timing and content is
necessary.

\paragraph*{Carrier phase}

Considering the reception of bits at the physical layer, knowledge
of the carrier phase at the receiver is crucial for successful reception
of phase modulated signals because the information is carried in the
phase variations of the signal, such that these offsets should be
minimized \cite{07:DigiCom:book}. Typically this is achieved during
the synchronization phase of packet reception, and thus existing capture
models have omitted phase offsets. However, there are two reasons
why this is not sufficient. First, in novel protocols exploiting packet
collisions, the synchronization during the preamble is not always
able to succeed. Second, there are other new applications of concurrent
transmissions that try to abandon the synchronization procedure. For
example, Pöpper et al.~\cite{11:Wireless-Manip:ESO} investigate
the possibility of manipulating individual message bits on the physical
layer, and conclude that carrier phase offsets are the major hindrance
to do so reliably.

\section{System Model}

\label{sec:Model}\global\long\def\ai{\alpha_{k}^{I}}
\global\long\def\aq{\alpha_{k}^{Q}}
\global\long\def\bi{\beta_{k}^{I}}
\global\long\def\bnq{\beta_{k-1}^{Q}}
\global\long\def\bq{\beta_{k}^{Q}}
\global\long\def\ki{k^{\prime}}
\global\long\def\bkni{\beta_{k-1}^{I}}
\global\long\def\bki{\beta_{k}^{I}}
\global\long\def\kq{k^{Q\prime}}
\global\long\def\bknq{\beta_{\kq-1}^{Q}}
\global\long\def\bkq{\beta_{\kq}^{Q}}
\global\long\def\ti{\tau_{i}}
\global\long\def\tuq{\underline{\tau^{Q}}}
\global\long\def\tu{\underline{\tau}}
\global\long\def\p{\varphi_{p}}
\global\long\def\ppi{\varphi_{p,i}}
\global\long\def\c{\varphi_{c}}
\global\long\def\cci{\varphi_{c,i}}
\global\long\def\pq{\varphi_{p}^{Q}}
\begin{figure*}
\makebox[1\textwidth]{%
\centering \includegraphics{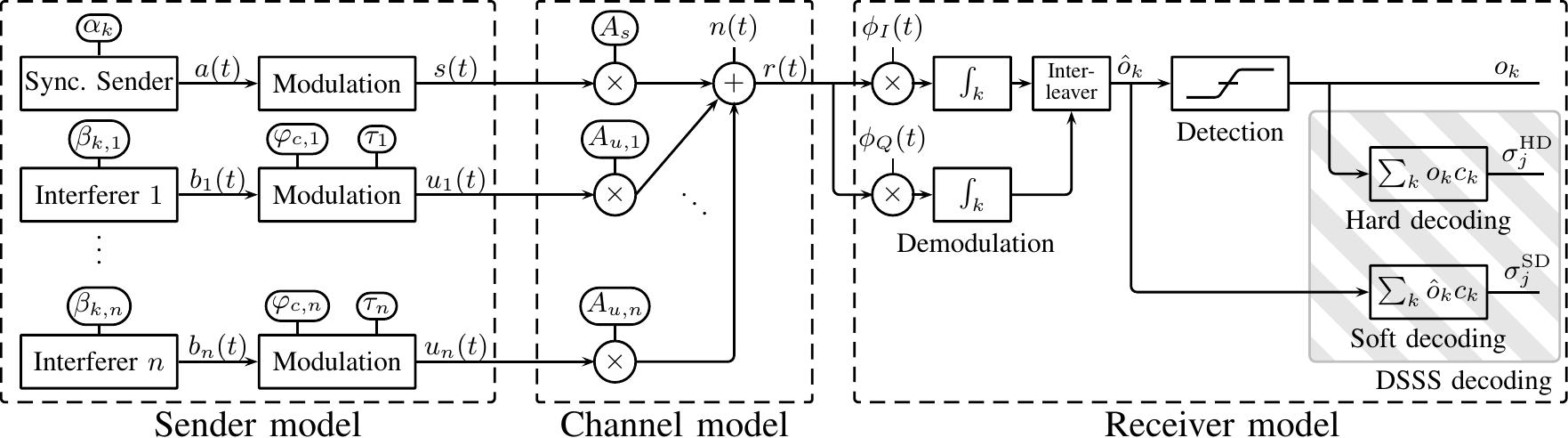}%
}

\caption{\label{fig:SystemModel}System model, its parameters are shown in
ovals (information bits $\alpha_{k},\beta_{k}$, carrier phase offset
$\varphi_{c}$, time offset $\tau$, signal amplitudes $A_{s},A_{u}$).
We consider one synchronized sender and $n$ interferers on a collision
channel that is the input to a receiver. Here, three channel coding
schemes are considered, \emph{(i)} uncoded, \emph{(ii)} DSSS with
hard decision decoding (HDD), and \emph{(iii)} DSSS with soft decision
decoding (SDD); resulting in different receiver paths.}
\end{figure*}
In this section, we discuss the system model underlying our analysis,
as shown in \figref{SystemModel}. It considers all factors from the
previous section. From a bird's eye view, the model consists of three
components: \emph{(i)} the sender model that modulates the physical
layer signals of $n+1$ transmitters, one fully synchronized signal
of interest (SoI) and $n$ interferers with possibly differing transmission
starting times and payloads; \emph{(ii)} the channel model with all
senders sharing a single collision channel that outputs a scaled superposition
of all signals (according to their corresponding power at the receiver),
and \emph{(iii)} the receiver model with three detection methods:
uncoded, DSSS with hard decision decoding (HDD), and DSSS with soft
decision decoding (SDD). In the following, we discuss each component
in detail. The notation used is collected in \tabref{Notation}.

\begin{table}[p]
\vspace{4cm}
\makebox[1\columnwidth]{%
\normalsize %
\begin{tabular}{l|l}
Symbol & Definition\tabularnewline
\hline 
$s(t)$ & MSK signal by the synchronized sender as defined in \eqref{MSKsig}\tabularnewline
$u_{i}(t)$ & MSK signal by interferer $i$, with possible offsets $\ti$ and $\cci$
(\eqref{ui})\tabularnewline
$r\left(t\right)$ & Resulting superposition of signals at the receiver (\eqref{r})\tabularnewline
$2T$ & Bit duration, e.g., $2T=\SI{1}{\micro\second}$ in IEEE 802.15.4\tabularnewline
$\omega_{c}=2\pi f_{c}$ & Angular speed of the carrier wave with frequency $f_{c}$\tabularnewline
$\omega_{p}=\frac{\pi}{2T}$ & Angular speed of baseband pulses (periodic by $4T$)\tabularnewline
$\ti$ & Time offset (positive shifts denote a starting delay)\tabularnewline
$\cci$ & Carrier phase offset in the passband of interferer $i$\tabularnewline
$\ppi=\omega_{p}\ti$ & Baseband pulse phase offset of interferer $i$, equivalent to time
offset\tabularnewline
$A_{s,}A_{u_{i}}$ & Signal amplitudes of $s\left(t\right)$, $u_{i}\left(t\right)$ at
the receiver\tabularnewline
$\Pi\left(t\right)$ & Unit pulse (step) function as defined in \eqref{Pulse}\tabularnewline
$a_{I},a_{Q}\left(t\right)$; $b_{I,}b_{Q}\left(t\right)$ & Information sequences consisting of unit pulses $\Pi(t)$ (\eqref{aIpulse})\tabularnewline
$\alpha_{k}^{I},\alpha_{k}^{Q}$; $\beta_{k}^{I},\beta_{k}^{Q}$ & Information bit $k$ of the synchronized sender and interferers\tabularnewline
$\phi_{I},\phi_{Q}\left(t\right)$ & Basis function of the MSK modulation to demodulate bits\tabularnewline
$\Lambda_{u}\left(k\right)$ & Contribution of signal $u\left(t\right)$ to the bit decision in bit
interval $k$ (\eqref{Lambda})\tabularnewline
$\hat{o}_{k}^{I}$ & Decision variable of the detector for bit $k$ of $I$ component\tabularnewline
$o_{k}^{I}$ & Detected bit of an uncoded transmission\tabularnewline
$\xi$ & Input symbol at the sender\tabularnewline
$c_{\xi,k}$  & Chipping sequence of symbol $\xi$ (see also \tabref{ChipSeqs})\tabularnewline
$\sigma_{j}^{HD},\sigma_{j}^{SD}$ & Detected symbol after DSSS decoding, the index $j$ \tabularnewline
 & ~~compensates that each symbol consists of 16 IQ pairs (see \eqref{DSSSCorr})\tabularnewline
$k^{\prime}=k-\left\lfloor \tau/2T\right\rfloor $ & Correction factor for the bits active in a decision interval\tabularnewline
$\underline{\tau}=\tau-2k^{\prime}T$ & Relative shift in a bit of interest $k$\tabularnewline
$k^{Q\prime}=k-\left\lfloor \left(\tau+T\right)/2T\right\rfloor $ & Correction factor for Q bits during I detection\tabularnewline
$k^{I\prime}=k-\left\lfloor \left(\tau-T\right)/2T\right\rfloor $ & Correction factor for I bits during Q detection\tabularnewline
$\underline{\tau^{Q}}=\tau+T-2k^{Q\prime}T$ & Relative shift in a bit of interest $k$ for the leaking Q-phase\tabularnewline
$\underline{\tau^{I}}=\tau-T-2k^{I\prime}T$ & Relative shift in a bit of interest $k$ for the leaking I-phase\tabularnewline
\end{tabular}%
}

\caption{\selectlanguage{american}%
\label{tab:Notation}\foreignlanguage{english}{Notation used in the
derivations.}\selectlanguage{english}%
}
\vspace{4cm}
\end{table}

\subsection{Sender Model}

In the first component, we modulate the physical signals of $n+1$
senders. We instantiate our model with the Minimum Shift Keying (MSK)
modulation, a widely used digital modulation with desirable properties,
and of special interest because of its use in the 2.4\,GHz PHY of
IEEE 802.15.4 \cite[$\S$6.5]{ieee802.15.4}, but we also discuss other
modulation schemes including O\nobreakdash-QPSK, QPSK, and BPSK.
For the signal representations, we follow the notation of Proakis
and Salehi \cite[$\S$4.3]{07:DigiCom:book}.

\subsubsection{Synchronized sender}

\begin{figure}
\includegraphics{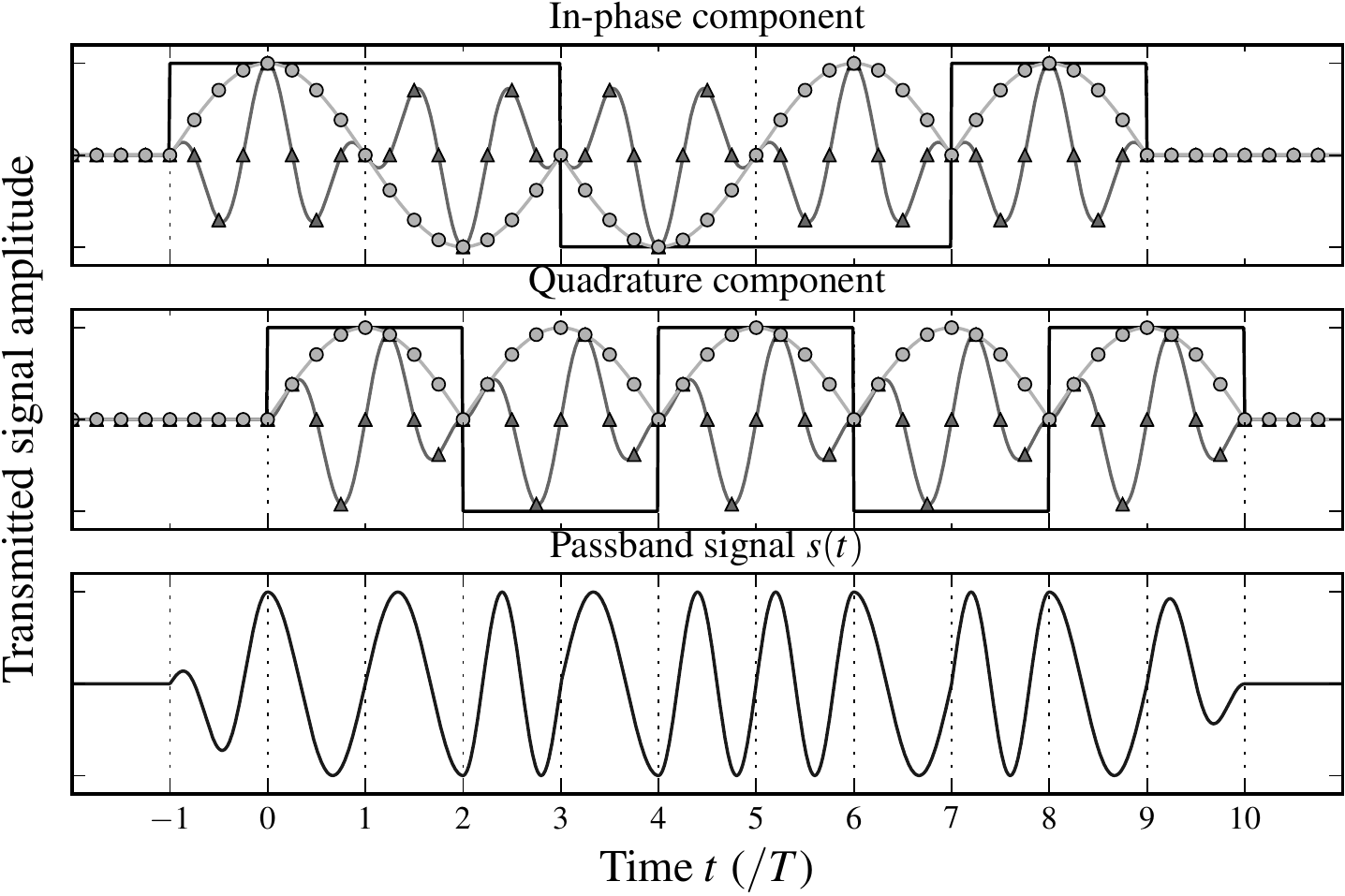}

\caption{\selectlanguage{american}%
\label{fig:MSKwave}\foreignlanguage{english}{MSK modulation example.
The modulated bit sequence is \texttt{1}\texttt{\textbf{1}}\texttt{1}\texttt{\textbf{0}}\texttt{0}\texttt{\textbf{1}}\texttt{0}\texttt{\textbf{0}}\texttt{1}\texttt{\textbf{1}}
(quadrature-phase bits are shown in boldface); it is multiplexed to
the IQ components (blue rectangular pulse trains), pulse-shaped with
half-sines (red sinusoids with $\circ$ markers) and modulated on
a carrier, resulting in the green waveform (with $\bigtriangleup$
markers). For the quadrature component, we observe an additional staggering
of $T$ (MSK can be viewed as Offset-QPSK with half-sine pulse shaping).
Both modulated IQ signals are added to result in the (real-valued)
passband time signal in the bottom figure.}\selectlanguage{english}%
}
\end{figure}

We assume that the receiver is fully synchronized to the SoI, i.e.,
the synchronization process has successfully acquired this signal
and all interferers have relative offsets to it. The signal is then
given by 
\begin{equation}
s\left(t\right)=a_{I}\left(t\right)\cos\left(\frac{\pi t}{2T}\right)\cos\omega_{c}t+a_{Q}\left(t\right)\sin\left(\frac{\pi t}{2T}\right)\sin\omega_{c}t.\label{eq:MSKsig}
\end{equation}
The signal consists of two components, the in- (I) and the quadrature-phase
(Q) components. Modulated onto each component are the information
signals (carrying the bits represented by $\alpha_{k}^{I},\alpha_{k}^{Q}\in\left\{ \pm1\right\} $)
given by
\begin{align}
a_{I}\left(t\right)= & \sum_{k=-\infty}^{\infty}\ai\Pi\left(\frac{t-2kT}{2T}\right)\label{eq:aIpulse}\\
a_{Q}\left(t\right)= & \sum_{k=-\infty}^{\infty}\aq\Pi\left(\frac{t-\left(2k+1\right)T}{2T}\right),\label{eq:aQpulse}
\end{align}
which represents a train of unit pulses $\Pi$ with duration $2T$,
the bit duration of the modulation (e.g., $2T=1$\,$\mu$s in IEEE
802.15.4). The unit pulses are defined by 
\begin{equation}
\Pi\left(t\right)=\begin{cases}
0 & \mbox{if }\left|t\right|>\frac{1}{2}\\
\frac{1}{2} & \mbox{if }\left|t\right|=\frac{1}{2}\\
1 & \mbox{if }\left|t\right|<\frac{1}{2}
\end{cases}\label{eq:Pulse}
\end{equation}
The information signals are staggered, i.e., the Q-phase information
signal is delayed by $T$ in $a_{Q}\left(t\right)$. These signals
are then shaped with half-sine pulses of duration $2T$, and modulated
onto a carrier with frequency $\omega_{c}/2\pi$ (e.g., 2.4--2.48\,GHz
in IEEE 802.15.4). In the following, we use the angular frequency
of baseband pulses $\omega_{p}=\pi/2T$, such that the first cosine
term in Eq.~(\ref{eq:MSKsig}) may be represented by $\cos\omega_{p}t$.
A graphical illustration of such an MSK-modulated signal is shown
in \figref{MSKwave}.

\subsubsection{(Unsynchronized) interferers}

In addition to the synchronized sender, we consider $n$ interferers
transmitting concurrently, using the same modulation. These signals
may not be synchronized to the receiver and each may carry its own
payload. This introduces three additional parameters that influence
the signal, the time offset $\tau_{i}$, the carrier phase offset
$\varphi_{c,i}$, and the information bits $\beta_{k,i}$. With a
positive $\tau_{i}$, an interfering signal arrives later at the receiver
than the synchronized signal. The signal at the receiver for interferer
$i$ is given by 
\begin{align}
u_{i}\left(t;\tau_{i},\varphi_{c,i}\right)= & \, b_{I,i}\left(t-\tau_{i}\right)\cos\omega_{p}\left(t-\tau_{i}\right)\cos\left(\omega_{c}t+\varphi_{c,i}\right)\nonumber \\
 & \hspace{-11bp}+b_{Q,i}\left(t-\tau_{i}\right)\sin\omega_{p}\left(t-\tau_{i}\right)\sin\left(\omega_{c}t+\varphi_{c,i}\right).\label{eq:ui}
\end{align}

We assume that the phase offsets $\varphi_{c,i}$ are constant for
the duration of a packet, i.e., there is no carrier frequency offset
during a transmission. In our experiments in \secref{Experiments},
we show that this assumption is reasonable because receiver implementations
are compensating for possible drifts. For convenience, we express
the pulse phase offset caused by $\tau_{i}$ as $\varphi_{p,i}=\omega_{p}\tau_{i}$.

\subsubsection{Other modulation schemes}

While our results are derived for the MSK modulation, it is possible
to adapt them to other variants of the phase shift keying (PSK) modulation.
We briefly describe the differences to major variants and highlight
how these affect the analysis. Further details on the relationship
between PSK modulation schemes can be found in Proakis and Salehi~\cite{07:DigiCom:book}
and Pasupathy \cite{79:MSK:ComMag}.

\subsubsection*{Offset QPSK}

O-QPSK with a half-sine pulse shape is identical to MSK \cite{05:GNUDeEn:TR}
and the results therefore also apply for this modulation. If O-QPSK
is used in combination with rectangular pulse shaping instead, the
signal is then given by
\[
s_{\mathrm{O-QPSK}}\left(t\right)=\frac{1}{\sqrt{2}}\left(a_{I}\left(t\right)\cos\omega_{c}t+a_{Q}\left(t\right)\sin\omega_{c}t\right).
\]
The altered pulse shape leads to the omission of the factor $\cos\omega_{p}t$
present in Eq.~(\ref{eq:MSKsig}), because the rectangular shaping
is already included in the information signal $a\left(t\right)$.
This leads to a simplification of our MSK results because pulse phase
offsets $\varphi_{p}$ that are caused by the time offset $\tau$
are not present.

\subsubsection*{\textit{Quadrature PSK}}

Considering \emph{QPSK}, the change from O-QPSK is the missing time
shift $T$ in the quadrature phase. This leads to a different information
signal for the Q phase, 
\[
a_{Q}^{\prime}\left(t\right)=\sum_{k=-\infty}^{\infty}\aq\Pi\left(\frac{t-2kT}{2T}\right).
\]
When adapting our results to QPSK, this affects the indices $k$ of
the colliding bits.

\subsubsection*{\textit{Binary PSK}}

This scheme considers only the in-phase components of QPSK, its signal
is given by 
\[
s_{\mathrm{BPSK}}\left(t\right)=\frac{1}{\sqrt{2}}a_{I}\left(t\right)\cos\omega_{c}t.
\]
This simplifies the derivations and results further, because there
is no contribution from the Q phase signal in collisions.

\subsection{Channel Model}

In our model, we use an additive collision channel. The relation for
the output signal is
\begin{equation}
r\left(t\right)=A_{s}\, s\left(t\right)+\sum_{i=1}^{n}A_{u,i}\, u_{i}\left(t;\tau_{i},\varphi_{c,i}\right)+n\left(t\right).\label{eq:r}
\end{equation}
Each signal is scaled by a positive, real-valued factor $A$, which
contains both, possible signal amplifications by the sender and path
loss effects that reduce the power at the receiver. In our evaluation,
we use the Signal to Interference Ratio (SIR) at the receiver, given
by $\mathrm{SIR}=A_{s}^{2}/\left(\sum_{i=1}^{n}A_{u,i}^{2}\right)$,
to characterize the power relationship of the interfering signals.
The contribution of all noise effects is accumulated in the linear
noise term $n\left(t\right)$; possible instantiations are a noiseless
channel or a white Gaussian noise channel.

\subsection{Receiver Model}

In the final component of the model, we feed the signals' superposition
$r\left(t\right)$ into an optimal receiver to discern the detected
bits. The signal is demodulated and fed into one of three detector
implementations: one for uncoded bits, and two variants of DSSS decoding.

\subsubsection{Demodulation}

Demodulation is performed for I and Q individually and the bits are
then interleaved. We limit our discussion to the I component for brevity.

We use the matched filter function $\phi_{I}\left(t\right)=\left(2/T\right)\cos\omega_{p}t\cos\omega_{c}t$
and low-pass filtering for downconversion and demodulation, which
is the optimal receiver for noiseless and Gaussian channels in the
sense that it minimizes the bit error probability \cite[§4.3]{07:DigiCom:book}.
The received signal $r\left(t\right)$ is multiplied by $\phi_{I}\left(t\right)$
and integrated for each bit period $k$ to form the decision variable

\begin{equation}
\hat{o}_{k}^{I}=\Lambda_{r}^{I}\left(k\right)=\int_{\left(2k-1\right)T}^{\left(2k+1\right)T}r\left(t\right)\phi_{I}\left(t\right)dt.\label{eq:Lambda}
\end{equation}
The resulting (real) value is called \emph{soft bit}. Because the
combination of the interferers in the received signal is linear, the
individual contributions can be divided into integrals for each signal:
\[
\hat{o}_{k}^{I}=\Lambda_{s}^{I}\left(k\right)+\sum_{i=1}^{n}\Lambda_{u_{i}}^{I}\left(k\right)+\Lambda_{n}^{I}\left(k\right).
\]
In our analytical evaluation in the following section, we derive closed-form
expressions for $\Lambda_{u_{i}}^{I}$ and $\Lambda_{u_{i}}^{Q}$
to analyze the receiver output after a signal collision.

We point out that this simplified model does not include receiver-side
techniques such as Automatic Gain Control (AGC) or phase tracking;
however, we conjecture that the reception performance is still comparable.
In fact, as our experiments in \secref{Experiments} show, this assumption
is justified and the simplified model is able to predict the reception
behavior of real-world receiver implementations with good accuracy.
We leave the investigation on the effects of these advanced techniques
to future work.

\subsubsection{Uncoded bit detection}

The detection operation for uncoded transmissions is slicing, essentially
a sign operation on the demodulation output, which results in binary
output $o_{k}\in\left\{ \pm1\right\} $. Thus, a bit of the SoI is
flipped if the contribution of the interferers changes the bit's sign.

\subsubsection{DSSS decoding}

For coded transmissions, the number of chips exceeds the bits in a
symbol, i.e., even if several chips are flipped it is still possible
to decode a symbol correctly. We consider $2^{b}$ symbols $\xi$
with chipping sequence $c_{\xi}$, each with a block length of $B$\,bit
(i.e., the number of chips). For example, we have $b=4$, $B=32$
in IEEE 802.15.4. 

We differentiate two modes of operation for the DSSS decoder, namely
hard decision decoding (HDD) and soft decision decoding (SDD) \cite{07:DigiCom:book}.

\subsubsection*{Hard decision decoding}

In \emph{HDD}, the decoder uses sliced (binary) values $o_{k}$ as
its input, and then chooses the symbol with the highest bit-wise cross-correlation
of all chipping sequences. In this way, HDD can be viewed as an additional
step that takes a group of uncoded bits with $B$ elements (from the
uncoded bit detection described above) to determine a symbol $\sigma_{j}^{\mathrm{HD}}$,
i.e., a group of $b$ bits. For HDD, the decoder is given by 
\begin{equation}
\sigma_{j}^{\mathrm{HD}}=\arg\max_{0\leq\xi<2^{b}}\left|\sum_{k=0}^{B-1}o_{jB+k}\: c_{\xi,k}\right|.\label{eq:DSSSCorr}
\end{equation}

\subsubsection*{Soft decision decoding}

In \emph{SDD}, the real-valued, unquantized demodulator output $\hat{o}_{k}$
(\emph{soft bits}) is used as decoder input directly, in contrast
to the binary values $o_{k}$ used in HDD. This is beneficial because
soft bits provide a measure of detection confidence and demodulation
quality, and thus adds weighting to the bits used in the cross-correlation.

\section{Mathematical Analysis}

\label{sec:Analysis}
\begin{table*}
\centering %
\makebox[1\textwidth]{%
\begin{tabular}{lll}
No offsets & $\Lambda_{u}^{I}\left(k\right)=A_{u}\bi$ & (1)\tabularnewline[2mm]
\hline 
\noalign{\vskip1mm}
Carrier phase offset $\varphi_{c}$  & $\Lambda_{u}^{I}\left(k\right)=A_{u}\left(\cos\varphi_{c}\bi-\frac{1}{\pi}\sin\varphi_{c}\left(\bnq-\bq\right)\right)$ & (2)\tabularnewline[2mm]
\hline 
\noalign{\vskip1mm}
Time offset $\tau$ & $\Lambda_{u}^{I}\left(k\right)=\frac{1}{2T}A_{u}\left(\cos\varphi_{p}\left(\tu\bkni+\left(2T-\tu\right)\bki\right)-\frac{2T}{\pi}\sin\varphi_{p}\left(\bkni-\bki\right)\right)$ & (3)\tabularnewline[2mm]
\hline 
\noalign{\vskip1mm}
Carrier phase + time offset & $\Lambda_{u}^{I}\left(k\right)=\frac{1}{2T}A_{u}\left\{ \cos\varphi_{c}\left(\cos\varphi_{p}\left(\tu\bkni+\left(2T-\tu\right)\bki\right)-\frac{2T}{\pi}\sin\varphi_{p}\left(\bkni-\bki\right)\right)\right.$ & (4)\tabularnewline[2mm]
\noalign{\vskip1mm}
 & $\hphantom{\Lambda_{u}^{I}\left(k\right).=iA_{u}\,}\left.-\sin\varphi_{c}\left(\sin\varphi_{p}\left(\tuq\bknq+\left(2T-\tuq\right)\bkq\right)+\frac{2T}{\pi}\cos\varphi_{p}\left(\bknq-\bkq\right)\right)\right\} $ & \tabularnewline[2mm]
\end{tabular}%
}

\caption{\label{tab:AnalysisResults}Analytical results: contributions of an
interfering signal to the demodulator output $\Lambda_{u}^{I}$ for
the $k$-th bit. The results present the relationship between in-
and quadrature phase bits sent by an interferer ($\beta^{I}$ and
$\beta^{Q}$) with time-adjusted bit indices $k^{\prime},k^{Q\prime}$;
it also considers the effects of carrier phase offsets $\varphi_{c}$
and time offsets ($\varphi_{p}$ and $\tau$). The corresponding notation
is introduced in Sections~\ref{sec:Model} and \ref{sec:Analysis}.}
\end{table*}
Based on the system model in \figref{SystemModel}, we analyze the
contributions of each interfering signal to the overall demodulator
output; the sum of these contributions is the decision variable of
bit detection. We first present the general case considering all system
parameters in \thmref{FullyUSync}. Subsequently, we illustrate its
interpretation using selected parameter combinations.
\begin{thm}
\label{thm:FullyUSync}For an interfering MSK signal $u\left(t\right)$
with offset parameters $\tau$ and $\varphi_{c}$, the contribution
to the demodulation output $\Lambda_{u}^{I}\left(k\right)$ is given
by Eq.~(4) in \tabref{AnalysisResults}.%
\footnote{We omit the subscript $i$ for clarity in the equations. The results
for the quadrature phase are given by the same equations when the
roles of I and Q are exchanged.%
}
\end{thm}
The proof of this theorem can be found in \apxref{Offset-Both-I}.
To provide a better understanding of the effects of the parameters,
we focus on selected parameter constellations and discuss the resulting
equations. Then we revisit \thmref{FullyUSync} and discuss the combination
of effects. \setcounter{subsubsection}{0}

\subsubsection{Synchronized signal}

In the simplest case both offsets, time and phase, are zero, i.e.,
the interfering signal is also fully synchronized to the receiver.
The result is given in Eq.~(1) in \tabref{AnalysisResults}. The
signal's contribution to the $k$-th bit is $\Lambda_{u}^{I}\left(k\right)=A_{u}\bi$.
The bit decision of bit $k$, i.e., the sign of the equation, is governed
by $\bi$. The magnitude of the contribution is controlled by the
amplitude of the signal $A_{u}$, and thus stronger signals lead to
a greater contribution to the decision variable $\hat{o}_{k}$. As
an example, consider two signals $s\left(t\right)$ and $u\left(t\right)$
that are both fully synchronized to the receiver. The detector output
of bit $k$ is then $A_{s}\ai+A_{u}\bi$. If both senders transmit
the same bit ($\ai=\bi$), then the signals interfere constructively
and push the decision variable further away from zero. If, on the
other hand, the bits are different, then the decision variable has
the sign of the stronger signal; this is the well-known power capture
effect for a single bit.
\begin{figure}[h]
\includegraphics{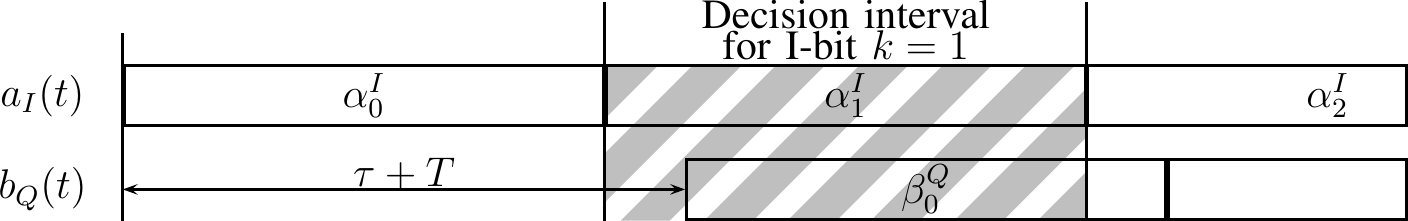}

\caption{\label{fig:IntervalPhi}Carrier phase offset $\varphi_{c}$: several
bits influence the bit decision on bit $k$ in a collision between
two signals. The carrier phase offsets lead to a leakage of the quadrature
phase, and because the Q-bits are staggered, there is an additional
shift of $T$ in the bit indices. The active bits in the decision
interval are highlighted.}
\end{figure}

\subsubsection{Carrier phase offset}

Next, we analyze the effect of carrier phase offsets when the signals
are fully time-synchronized ($\tau=0$), as shown in \figref{IntervalPhi}.
The result is given in Eq.~(2). We observe two effects of the carrier
phase offset. First, the bit contribution of $\bi$ is scaled by $\cos\varphi_{c}\leq1$,
which leads to reduced absolute values (and thus a smaller contribution
to the decision variable) and potentially causes the bit $\bi$ to
flip for $\varphi_{c}\in\left(\frac{\pi}{2},\frac{3\pi}{2}\right)$.
Second, the quadrature phase starts to leak into the decision variable
and thus two additional bits $\beta_{k-1}^{Q},\beta_{k}^{Q}$ influence
the outcome. This contribution, however, is scaled by $\pi^{-1}\sin\varphi_{c}$,
and only appears when the two Q bits are alternating during the integration
interval. In essence, uncontrolled carrier phase offsets may lead
to unpredictable bits in the detector output because of carrier phase
offset induced bit flips.
\begin{figure}
\centering \includegraphics{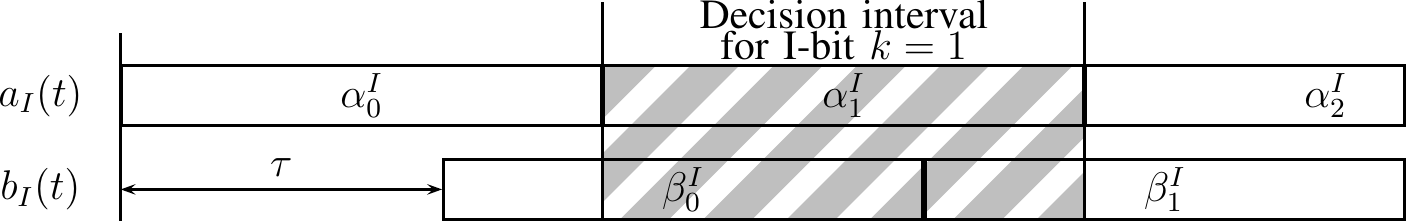}

\caption{\label{fig:IntervalTime}Example of a time offset $\tau$ during detection:
three bits influence the bit decision on the second bit in a collision
between two signals. The active bits in the decision interval are
highlighted (the synchronized sender's bit $\alpha_{1}^{I}$ and interferer's
in-phase bits $\beta_{0}^{I}$ and $\beta_{1}^{I}$).}
\end{figure}

\subsubsection{Time offset}

If the signals are phase-matched but shifted in time, the detector
output is given by Eq.~(3). We make three observations here. The
bit index $k$ needs to be adjusted because bits may be time-shifted
into the integration interval, see \figref{IntervalTime}; the new
index is given by $k^{\prime}=k-\left\lfloor \tau/2T\right\rfloor $,
with $\left\lfloor \cdot\right\rfloor $ denoting the floor function.
We call these \emph{active bits} because they contribute to the bit
decision. These bits overlap partially or fully, and their active
time duration is $\tu=\tau-2\left\lfloor \tau/2T\right\rfloor T$,
the underscore signifies that its value is confined to the interval
$\left[0;2T\right)$. However, these bits do not contribute to the
decision directly but are scaled by $\cos\varphi_{p}$, which is caused
by the half-sine pulse shaping of MSK. This scaling means that bit
contributions are diminished and may be flipped by certain time offsets.
Finally, a term scaled by $\pi^{-1}$ is introduced that is only present
when bits are alternating. However, these bits are the same in-phase
bits $\bkni,\bki$, the Q phase does not leak in this setting.

\subsubsection{Both offsets}

\begin{figure}
\centering \includegraphics{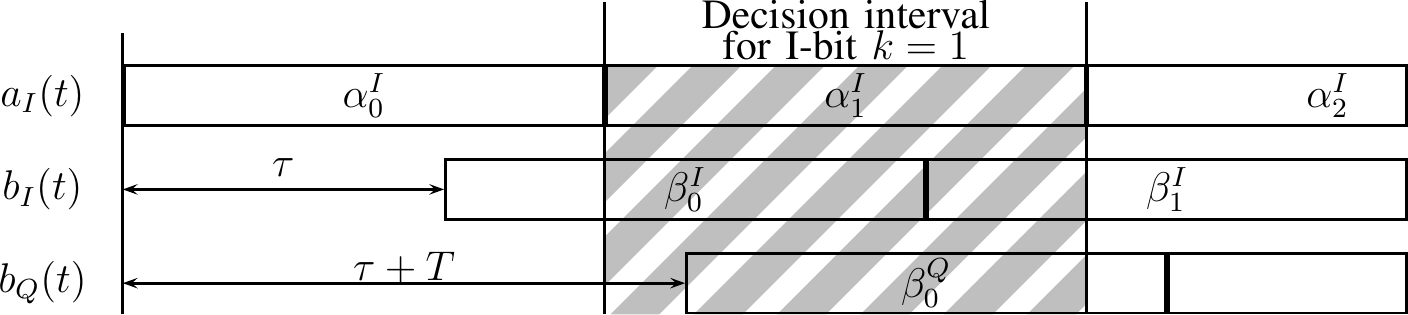}

\caption{\label{fig:IntervalBoth}Example of time and phase offsets combined:
the decision on the second bit ($k=1$) is influenced by four bits
in this example (the synchronized sender's bit $\alpha_{1}^{I}$,
the interferer's in-phase bits $\beta_{0}^{I}$ and $\beta_{1}^{I}$
due to time shifts, and quadrature bit $\beta_{0}^{Q}$ from carrier
phase offsets).}
\end{figure}
Finally, when both offsets are present as in \thmref{FullyUSync},
we can interpret the result as a combination of the above effects.
A graphical illustration of the active bits is shown in \figref{IntervalBoth}.
Due to the staggering of bits (the Q bits are delayed by $T$), the
indices of leaking bits of the Q phase also need to be adjusted, the
new index is $k^{Q\prime}=k-\left\lfloor \left(\tau+T\right)/2T\right\rfloor $,
and the active time interval $\underline{\tau^{Q}}$ is derived similarly
to above.

In summary, we observe that the contribution of the interfering signal
is complex and that $\varphi_{c}$ and $\varphi_{p}$ can potentially
flip the original bits $\bi$. This should be bad news for collision-aware
protocols that use identical payload to achieve constructive interference
(e.g., SCIF \cite{13:ScaleFlood:ToN}): these bits can flip easily
and then generate destructive interference. However, coding helps
to alleviate these negative effects as we will see in the next section.

\section{Parameter Space Exploration}

\label{sec:Evaluation}Equipped with the closed-form analytical model
of the bit-wise receiver outputs, we systematically explore the parameter
space of the reception of concurrent transmissions in detail.

\subsection{Methodology}

\begin{table*}

\begin{minipage}{1\linewidth}%
\centering\small\setlength{\tabcolsep}{3pt}%
\begin{tabular}{ccGcGcGcGcGcGcGcGcGcGcGcGcGcGcGcGc}
\hline 
\noalign{\vskip\doublerulesep}
Symbol $\xi$ & Bits & \multicolumn{32}{c}{Chipping sequence bits%
\footnote{The IQ chips are shown interleaved, dark background denotes in-phase
chips.%
}%
\footnote{The sequences are shifted cyclically by four chips, \textbf{bold chips}
are the first chips of symbol 0 (or 8) for reference.%
} ($c_{\xi,0},\ldots,c_{\xi,31}$)}\tabularnewline[\doublerulesep]
\hline 
0 & \hphantom{~~}0000~~ & \textbf{1} & \textbf{1} & \textbf{0} & \textbf{1} & 1 & 0 & 0 & 1 & 1 & 1 & 0 & 0 & 0 & 0 & 1 & 1 & 0 & 1 & 0 & 1 & 0 & 0 & 1 & 0 & 0 & 0 & 1 & 0 & 1 & 1 & 1 & 0\tabularnewline
1 & 0001 & 1 & 1 & 1 & 0 & \textbf{1} & \textbf{1} & \textbf{0} & \textbf{1} & 1 & 0 & 0 & 1 & 1 & 1 & 0 & 0 & 0 & 0 & 1 & 1 & 0 & 1 & 0 & 1 & 0 & 0 & 1 & 0 & 0 & 0 & 1 & 0\tabularnewline
2 & 0010 & 0 & 0 & 1 & 0 & 1 & 1 & 1 & 0 & \textbf{1} & \textbf{1} & \textbf{0} & \textbf{1} & 1 & 0 & 0 & 1 & 1 & 1 & 0 & 0 & 0 & 0 & 1 & 1 & 0 & 1 & 0 & 1 & 0 & 0 & 1 & 0\tabularnewline
3 & 0011 & 0 & 0 & 1 & 0 & 0 & 0 & 1 & 0 & 1 & 1 & 1 & 0 & \textbf{1} & \textbf{1} & \textbf{0} & \textbf{1} & 1 & 0 & 0 & 1 & 1 & 1 & 0 & 0 & 0 & 0 & 1 & 1 & 0 & 1 & 0 & 1\tabularnewline
4 & 0100 & 0 & 1 & 0 & 1 & 0 & 0 & 1 & 0 & 0 & 0 & 1 & 0 & 1 & 1 & 1 & 0 & \textbf{1} & \textbf{1} & \textbf{0} & \textbf{1} & 1 & 0 & 0 & 1 & 1 & 1 & 0 & 0 & 0 & 0 & 1 & 1\tabularnewline
5 & 0101 & 0 & 0 & 1 & 1 & 0 & 1 & 0 & 1 & 0 & 0 & 1 & 0 & 0 & 0 & 1 & 0 & 1 & 1 & 1 & 0 & \textbf{1} & \textbf{1} & \textbf{0} & \textbf{1} & 1 & 0 & 0 & 1 & 1 & 1 & 0 & 0\tabularnewline
6 & 0110 & 1 & 1 & 0 & 0 & 0 & 0 & 1 & 1 & 0 & 1 & 0 & 1 & 0 & 0 & 1 & 0 & 0 & 0 & 1 & 0 & 1 & 1 & 1 & 0 & \textbf{1} & \textbf{1} & \textbf{0} & \textbf{1} & 1 & 0 & 0 & 1\tabularnewline
7 & 0111 & 1 & 0 & 0 & 1 & 1 & 1 & 0 & 0 & 0 & 0 & 1 & 1 & 0 & 1 & 0 & 1 & 0 & 0 & 1 & 0 & 0 & 0 & 1 & 0 & 1 & 1 & 1 & 0 & \textbf{1} & \textbf{1} & \textbf{0} & \textbf{1}\tabularnewline
\hline 
\hphantom{$^{c}$}8%
\footnote{The second half of the chipping sequences are equal to the first except
that quadrature bits are inverted.%
} & 1000 & \textbf{1} & \textbf{0} & \textbf{0} & \textbf{0} & 1 & 1 & 0 & 0 & 1 & 0 & 0 & 1 & 0 & 1 & 1 & 0 & 0 & 0 & 0 & 0 & 0 & 1 & 1 & 1 & 0 & 1 & 1 & 1 & 1 & 0 & 1 & 1\tabularnewline
9 & 1001 & 1 & 0 & 1 & 1 & \textbf{1} & \textbf{0} & \textbf{0} & \textbf{0} & 1 & 1 & 0 & 0 & 1 & 0 & 0 & 1 & 0 & 1 & 1 & 0 & 0 & 0 & 0 & 0 & 0 & 1 & 1 & 1 & 0 & 1 & 1 & 1\tabularnewline
10 & 1010 & 0 & 1 & 1 & 1 & 1 & 0 & 1 & 1 & \textbf{1} & \textbf{0} & \textbf{0} & \textbf{0} & 1 & 1 & 0 & 0 & 1 & 0 & 0 & 1 & 0 & 1 & 1 & 0 & 0 & 0 & 0 & 0 & 0 & 1 & 1 & 1\tabularnewline
11 & 1011 & 0 & 1 & 1 & 1 & 0 & 1 & 1 & 1 & 1 & 0 & 1 & 1 & \textbf{1} & \textbf{0} & \textbf{0} & \textbf{0} & 1 & 1 & 0 & 0 & 1 & 0 & 0 & 1 & 0 & 1 & 1 & 0 & 0 & 0 & 0 & 0\tabularnewline
12 & 1100 & 0 & 0 & 0 & 0 & 0 & 1 & 1 & 1 & 0 & 1 & 1 & 1 & 1 & 0 & 1 & 1 & \textbf{1} & \textbf{0} & \textbf{0} & \textbf{0} & 1 & 1 & 0 & 0 & 1 & 0 & 0 & 1 & 0 & 1 & 1 & 0\tabularnewline
13 & 1101 & 0 & 1 & 1 & 0 & 0 & 0 & 0 & 0 & 0 & 1 & 1 & 1 & 0 & 1 & 1 & 1 & 1 & 0 & 1 & 1 & \textbf{1} & \textbf{0} & \textbf{0} & \textbf{0} & 1 & 1 & 0 & 0 & 1 & 0 & 0 & 1\tabularnewline
14 & 1110 & 1 & 0 & 0 & 1 & 0 & 1 & 1 & 0 & 0 & 0 & 0 & 0 & 0 & 1 & 1 & 1 & 0 & 1 & 1 & 1 & 1 & 0 & 1 & 1 & \textbf{1} & \textbf{0} & \textbf{0} & \textbf{0} & 1 & 1 & 0 & 0\tabularnewline
15 & 1111 & 1 & 1 & 0 & 0 & 1 & 0 & 0 & 1 & 0 & 1 & 1 & 0 & 0 & 0 & 0 & 0 & 0 & 1 & 1 & 1 & 0 & 1 & 1 & 1 & 1 & 0 & 1 & 1 & \textbf{1} & \textbf{0} & \textbf{0} & \textbf{0}\tabularnewline
\hline 
\end{tabular}%
\end{minipage}

\caption{Chipping sequences used in the 2.4\,GHz PHY of IEEE 802.15.4.}\label{tab:ChipSeqs}

\end{table*}
In order to numerically study the transmission reception success under
interference we perform so-called Monte Carlo simulations (see Jain
\cite{JainPerf}); that means we do time-static simulations of independent
packet transmissions in which we randomly vary the analytical model's
parameters to investigate their influence on performance parameters
such as packet reception ratio, bit and symbol error rate. Conceptually,
the simulator is just a software version of the mathematical model
(written in Python) applied to a whole packet; it is not meant to
validate the model but to experiment with randomly chosen values for
the model parameters and to provide more insights on the success probability
of concurrent transmissions. The simulation code is available for
download at \url{http://disco.cs.uni-kl.de/content/collisions}; there,
the interested reader can also find an interactive visualization of
the model.

For most experiments, the time offset between sender and interferer
is fixed and is our primary factor in the numerical analysis, i.e.,
in the plots we show the reception performance depending on the time
offset. The other parameters of the model are treated as secondary
factors and are randomly varied. Generating 1,000 independent packet
transmissions for each data point in the presented graphs thus represents
the secondary factors' average contribution to the reception success.
We provide more details on the choices for the model's parameters
for sender and channel in the following.

\subsubsection{Sender model}

For ease of presentation, we mainly consider the presence of one synchronized
sender and one interferer; we denote these parties as $\mathcal{S}$
and $\mathcal{I}$ with signals $s\left(t\right)$ and $u\left(t\right)$,
respectively. In \secref{N-Interferers}, we consider the $n$ interferer
case separately. We analyze the reception performance of groups of
associated bits, or packets; in this case, a single bit error leads
to a packet drop. The packet reception ratio (PRR) is the fraction
of packets that arrive without errors divided by the total number
of packets. We use packets with a length of 64\,bit. We consider
two categories of colliding packets, either with independent ($\mathcal{S}$
and $\mathcal{I}$ trying to exploit spatial reuse) or identical content
($\alpha_{k}=\beta_{k}$, as it is the case for collision-aware flooding
protocols). The bits to send are chosen in the following manner: for
uncoded transmissions, $\alpha_{k}$ is drawn bitwise i.i.d.~from
a Bernoulli distribution over $\left\{ -1,1\right\} $, and either
the same procedure is performed for $\beta_{k}$ (independent packets)
or simply copied over from $\alpha_{k}$ (identical packets). For
coded packets, we draw symbols i.i.d.~uniform random from $\left\{ 0,\ldots,15\right\} $
and spread these symbols according to the chipping sequences defined
by the IEEE 802.15.4 standard \cite[$\S$6.5]{ieee802.15.4}. This
means that 4\,bit groups are first spread to 32\,bit chipping sequences
before they are transmitted in $\alpha_{k},\beta_{k}$. The chipping
sequences are given in \tabref{ChipSeqs}. Note that for symbols \texttt{1}--\texttt{7},
the chipping sequences are shifted versions of the symbol \texttt{0},
while for the other half (symbols \texttt{8}--\texttt{15}), the quadrature-phase
bits are inverted. 

In accordance to the literature \cite{WC:Rappaport}, as the carrier
phase offset is hard to control because of oscillators drifts and
other phase changes during transmission, we draw $\varphi_{c}$ i.i.d.~uniform
randomly from $\left[0;2\pi\right)$ for each packet unless stated
otherwise. On the other hand, we use the same time offset $\tau$
for all packets because experimental work shows that this timing can
be precisely controlled. For example, Glossy \cite{Glossy} achieves
a timing precision of 500\,ns over 8 hops with 96\,\% probability,
and Wang et al.~\cite{Triggercast} report a 95\,\% percentile time
synchronization error of at most 250\,ns. For our simulations, we
used 1,000 packets for each value of $\tau$.

\subsubsection{Channel model}

To concentrate on the impact of signal interference, we consider a
noiseless channel. This is a well-accepted assumption when both signals
are significantly above noise floor level \cite[$\S$8]{PoiselJamming}.
We set $A_{s}=1$ and $A_{u}=\mathrm{SIR}^{-\frac{1}{2}}$.

\subsection{Reception of the Synchronized Signal of Interest}

\begin{figure*}
\centering \subfloat[\label{fig:Eval-Indep-Uncoded}Uncoded transmissions.]{\includegraphics{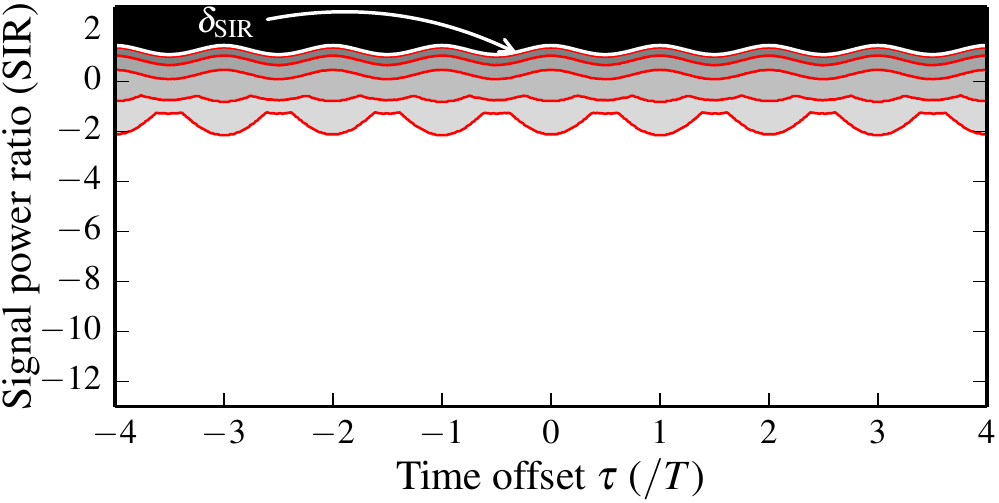}~\includegraphics{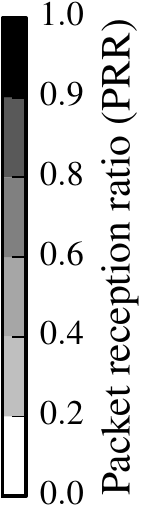}

}

\subfloat[\label{fig:Eval-Indep-HDD}DSSS with hard decision decoding.]{\includegraphics{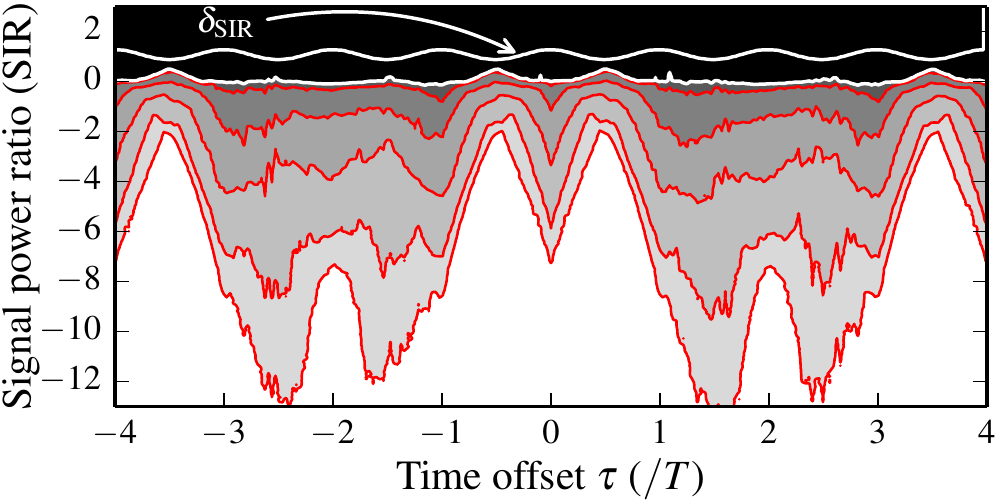}~\includegraphics{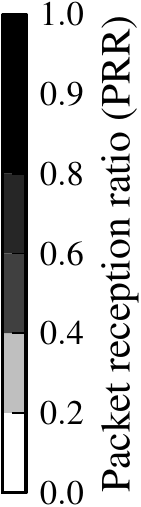}

}

\subfloat[\label{fig:Eval-Indep-SDD}DSSS with soft decision decoding.]{\includegraphics{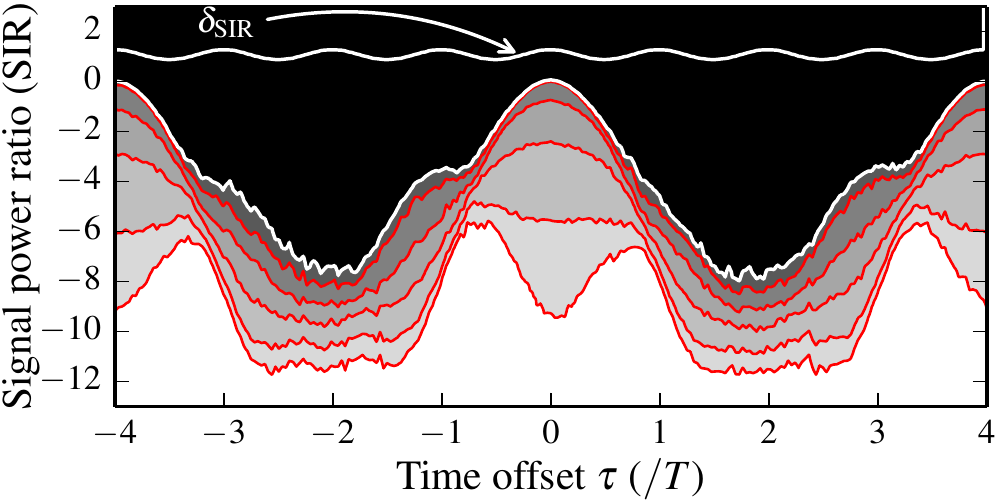}~\includegraphics{figs/1col/figs/ser_contour_AsAu/pdf/cb}

}

\caption{\label{fig:Eval-Indep}The capture threshold for two colliding packets
with \emph{independent} payload, varying with the signals' power ratio
SIR and time offset ($\tau=0$ indicates that the signals overlap
fully). For the uncoded case, the threshold $\delta_{\mathrm{SIR}}$
is nearly constant across all time offsets and represents the classical
capture threshold (thus, for reference, it is drawn in all figures).
For HDD, the threshold is nearly constant, but 1\,dB lower. Additionally,
there is a wide transitional region with non-zero PRRs. Finally, for
SDD the threshold is very sensitive to signal timing, we observe a
variation of 6--8\,dB with periodical time shifts of $2T$.}
\end{figure*}
\begin{figure}
\centering \subfloat[Uncoded transmissions.]{\includegraphics{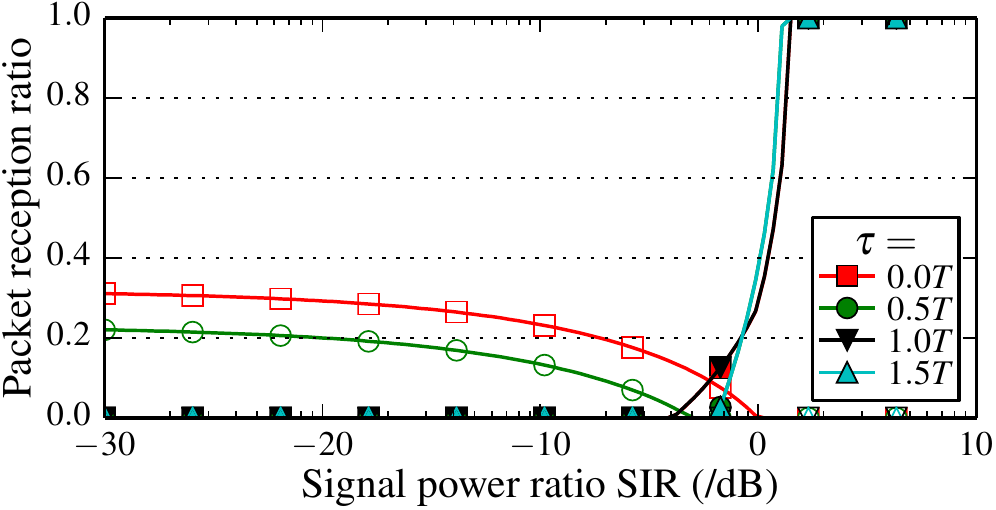}

}

\subfloat[DSSS with hard decision decoding (HDD).]{\includegraphics{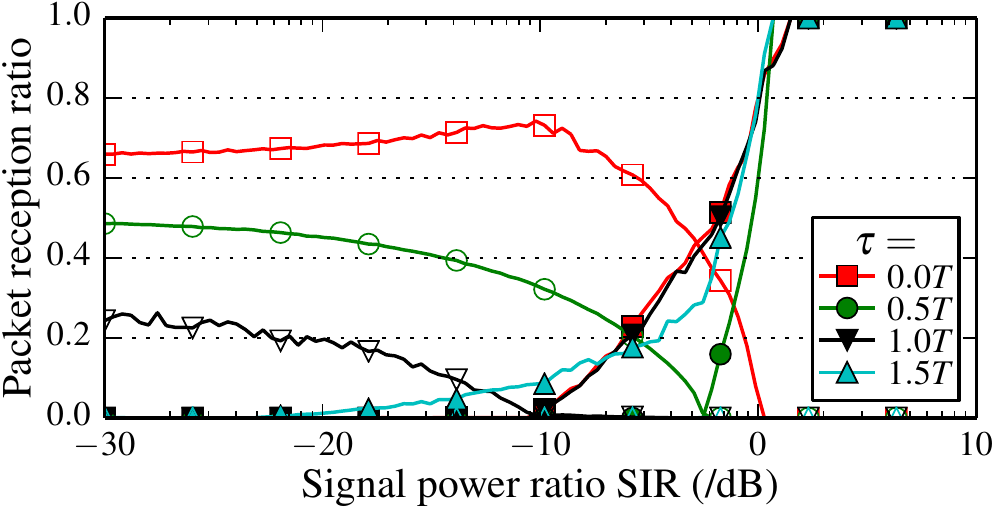}}

\subfloat[DSSS with soft decision decoding (SDD).]{\includegraphics{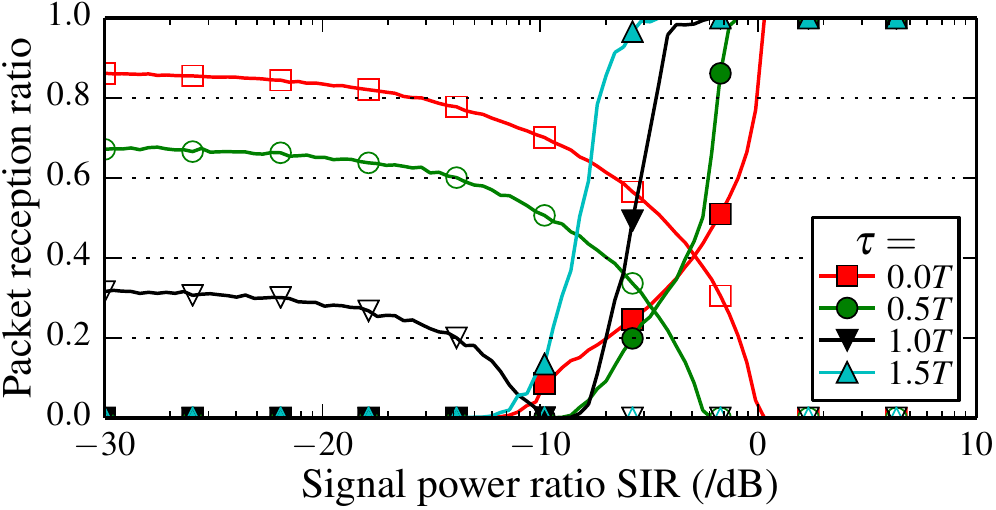}

}

\caption{\label{fig:Power-Independent}Effect of signal to interference ratio
SIR on the PRR for \emph{independent} payload. Filled markers represent
the reception of the synchronized sender's packets, empty markers
represent the reception of the interferer's packets. (a)~For uncoded
transmissions and in the negative SIR regime, the reception of the
interferer's packets is poor (max.~30\,\% PRR) even with perfect
time synchronization ($\tau=0$). The synchronized sender requires
a positive SIR for a high PRR independent of the interferers timing,
and the transitional region is narrow. (b)~For HDD, the interferer's
packets have an increasing chance of reception, strongly depending
on the timing. For the synchronized sender, the transitional region
is widened. (c)~This behavior is even more pronounced for the SDD
case, with a PRR of the interferer of over 85\,\% and the SIR transitional
zone has a width of 10\,dB.}
\end{figure}

\subsubsection{Capture threshold under independent payload}

In our first case study, we consider the transmission of independent
payload. This situation occurs, e.g., when two uncoordinated senders
detect a clear channel, transmit, and the packets collide at the receiver.
Our metric of interest is the PRR of the SoI, i.e., we observe the
probability to overcome the collision. The results for three classes
of receivers are shown in \figref{Eval-Indep} and \figref{Power-Independent}.

\textbf{Uncoded transmissions.} From \figref{Eval-Indep-Uncoded},
we observe that the capture threshold is a good model to describe
the PRR of interfering, uncoded transmissions. If the SoI is stronger
by a threshold $\delta_{\mathrm{SIR}}$ of 2\,dB, all its packets
are received.%
\footnote{For the numerical values of $\delta_{\mathrm{SIR}}$ shown in the
figures, we used a PRR threshold of 90\,\%.%
} This behavior persists for all choices of $\tau$, i.e., packet reception
is independent from the properties of the interfering signal (we only
see a minor periodic effect). Below the threshold, there is a narrow
transitional region with non-zero PRR. Under uncoded transmissions,
our model is able to recover the classical capture threshold for MSK
and is in accordance to experimental results in the literature \cite{10:154CaptureModel:ISWPC,ExpConcur}.

\textbf{Hard decision decoding.} When considering HDD (\figref{Eval-Indep-HDD}),
we note that the threshold abstraction is still valid and the performance
improvement of coding is only 1\,dB (the coding gain is canceled
when the same chipping sequences are used). In the transitional region,
there is a wider parameter range that results in non-zero PRRs, e.g.,
when $\tau$ is close to integer values (and thus $\cos\varphi_{p}\approx0$),
we observe a better PRR for $\mathcal{S}$. These results show that
coding with HDD yields only limited benefits if all senders use identical
chipping sequences.

\textbf{Soft decision decoding.} Finally, for SDD we observe a strong
dependence between PRR and time offset (\figref{Eval-Indep-SDD}).
Only for positions without chipping sequence shifts ($\tau=0$, and
because of the way IEEE 802.15.4 sequences are chosen%
\footnote{See \tabref{ChipSeqs}. The chipping sequences are not independently
chosen, they constitute shifted versions of a single generator sequence
with shifts of 4 IQ bits.%
}, $\tau=4kT$, $k\in\mathbb{Z}$) the performance is comparable to
the HDD case. For different time shifts, we can achieve a 6--8\,dB
coding gain despite the use of identical chipping sequences; especially
for offsets $\tau=4kT+2T$, we can achieve a clear coding gain. The
reason is that soft bits contain additional information on the detection
confidence, which helps to improve the detection performance in the
cross-correlation.

This insight suggests that two senders may benefit from coding even
when using independent payloads, provided that they time their collisions
precisely. This may help to increase the number of opportunities for
concurrent transmissions, i.e., interfering nodes can be much closer
to a receiver and still achieve the same PRR performance. In other
words, a \emph{constant} capture threshold is too conservative when
collision timing can be precisely controlled, because the performance
of SDD is very sensitive to time offsets.

\subsubsection{Capture threshold under identical payload}

\begin{figure*}
\centering \subfloat[\label{fig:Eval-Id-Uncoded}Uncoded transmissions.]{\includegraphics{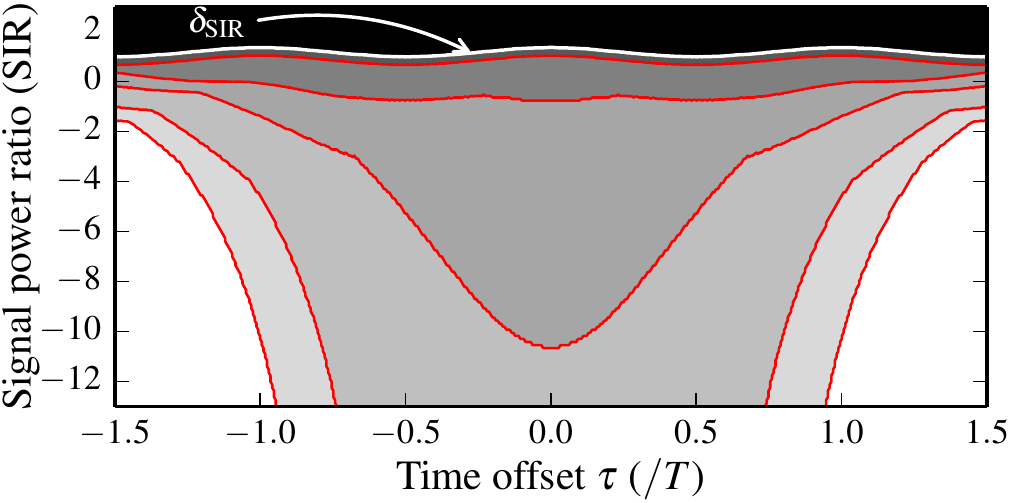}~\includegraphics{figs/1col/figs/ser_contour_AsAu/pdf/cb}

}

\subfloat[\label{fig:Eval-Id-HDD}DSSS with hard decision decoding.]{\includegraphics{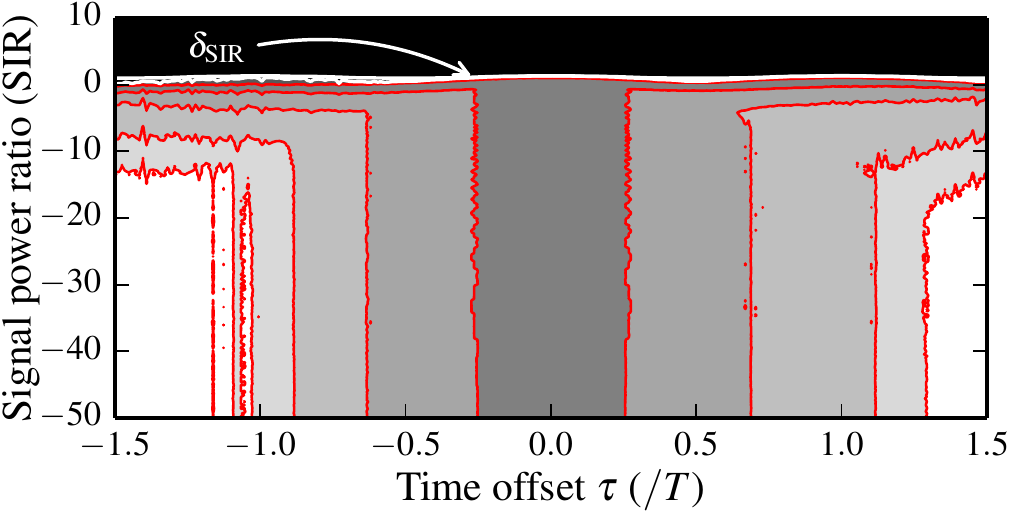}~\includegraphics{figs/1col/figs/ser_contour_AsAu/pdf/cb}

}

\subfloat[\label{fig:Eval-Id-SDD}DSSS with soft decision decoding.]{\includegraphics{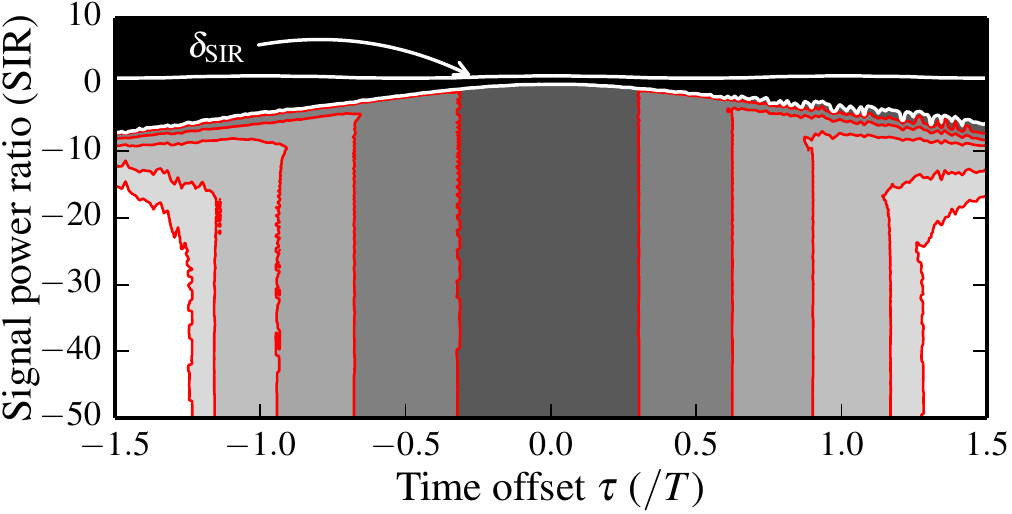}~\includegraphics{figs/1col/figs/ser_contour_AsAu/pdf/cb}

}

\caption{\label{fig:Eval-Id}The capture threshold for colliding packets with
\emph{identical} content depending on the power ratio SIR and the
time offset $\tau$. In all three figures, we show the threshold $\delta_{\mathrm{SIR}}$
for identical and uncoded payload as reference. (a) In the uncoded
case, the PRR is non-zero in the transitional range, but packet loss
is still likely with PRRs of 20--30\,\%. For coded transmissions,
we observe a central area that enables high PRR values (up to 70\,\%
in (b) and approximately 90\,\% in (c)).}
\end{figure*}
\begin{figure}
\centering \subfloat[Uncoded transmissions.]{\includegraphics{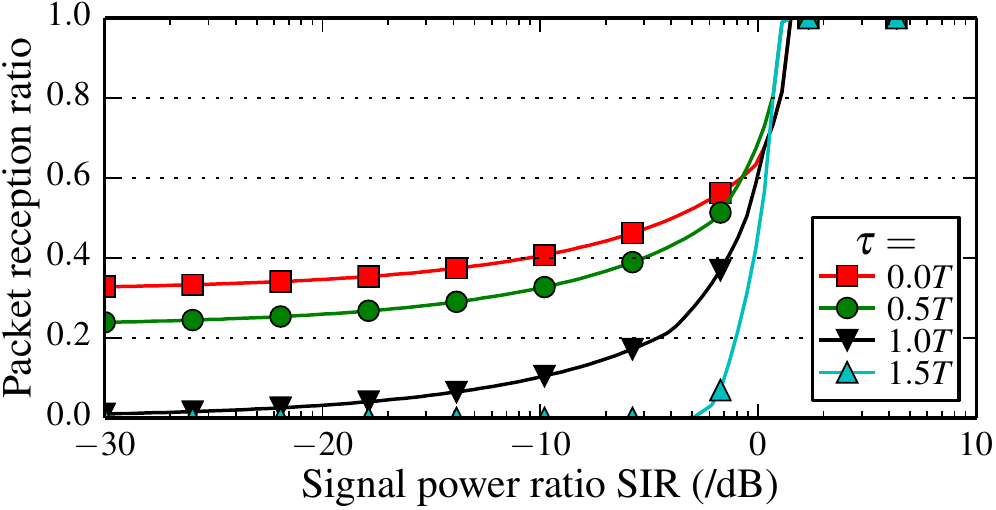}

}

\subfloat[DSSS with hard decision decoding (HDD).]{\includegraphics{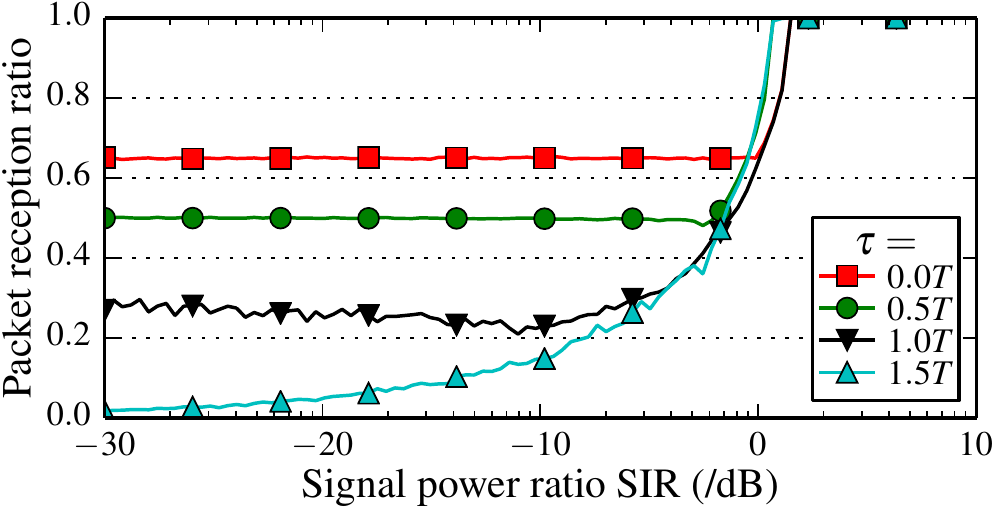}

}

\subfloat[DSSS with soft decision decoding (SDD).]{\includegraphics{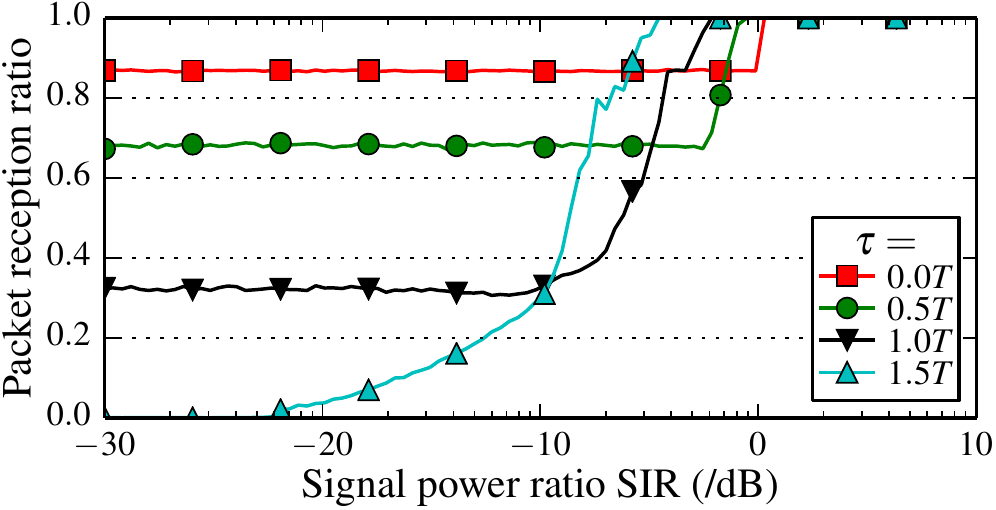}

}

\caption{\label{fig:Power-Identical}Effect of signal to interference ratio
SIR on the PRR for \emph{identical} payload. Because both use the
same payload, only filled markers are present in contrast to \figref{Power-Independent}.
(a)~For the uncoded case, the PRR varies from approx.~30\,\% to
100\,\% and is highly dependent on the SIR. (b)~For the HDD case,
the PRR takes two values that are more stable across the SIR range:
in the negative SIR regime the PRR is around 65\,\%, and 100\,\%
in the positive SIR regime. (c)~For the SDD case, the PRR for negative
SIR increases to over 85\,\% with perfect time synchronization, independent
of the SIR.}
\end{figure}
When considering the collisions of identical packets, we observe very
different results (\figref{Eval-Id} and \figref{Power-Identical}):
a good reception performance is possible despite even a negative SIR.%
\footnote{We note that with increasing time offsets $\tau$ the PRR performance
approaches the results for independent payloads.%
}

\textbf{Uncoded transmissions.} For uncoded transmissions, the PRR
performance is shown in \figref{Eval-Id-Uncoded}. While in this case
the threshold for a PRR of 100\,\% is still equal to the independent
payload case, substantially more packets are received in the transitional
region with time shifts less than $\pm0.75T$. However, PRRs around
30\,\% are usually not sufficient to boost the performance of network
protocols. The reason for this limited performance is the carrier
phase offset $\varphi_{c}$: with negative SIR, the interfering signal
dominates the bit decision at the receiver, and with larger offsets
$\varphi_{c}\in\left(\frac{\pi}{2};\frac{3\pi}{2}\right)$, the term
$\cos\varphi_{c}$ changes its sign and flips all subsequent bits.
In this sense, the literature conjecture that constructive interference
is the reason for the good performance of flooding protocols \cite{13:ScaleFlood:ToN,Triggercast}
is only valid if the receiver is synchronized to the strongest signal
and if the phase offset $\varphi_{c}$ can be neglected. However,
because the collisions start during the preamble when using such protocols,
successful synchronization cannot be ensured. Therefore, there must
be another mechanism that recovers flipped bits.

\textbf{Hard decision decoding.} The reception performance of coded
messages provides a hint in this direction (\figref{Eval-Id-HDD}).
We observe a corridor of $\tau$ values ($\tau=\pm0.2T$ or 100\,ns
in IEEE 802.15.4) that has a PRR of 60--80\,\% in the center (note
the larger SIR scale on the y-axis). When two signals with identical
payload collide with a small time offset, a reception is still possible
even if the interfering signal is far stronger. This suggests that
the interfering signal is received instead of the SoI, and that coding
helps to overcome bit flips of $\beta_{k}$ induced by the carrier
phase. The explanation is a property of Eq.~(\ref{eq:DSSSCorr}):
even if all bits are flipped by $\cos\varphi_{c}$, the (absolute)
correlation is still maximal for the correct chipping sequence. This
shows that DSSS used in IEEE 802.15.4 is a key factor to make the
collision-aware protocols work. 

\textbf{Soft decision decoding.} The experimentally observed performance
in the literature is even superior to \figref{Eval-Id-HDD} \cite{10:AMAC:SenSys,Glossy,13:ScaleFlood:ToN}.
Taking SDD into account, this gap is closed (\figref{Eval-Id-SDD}).
There is a strong center region for $\tau\leq\pm0.3T$, or 150\,ns
in 802.15.4, with a PRR of approximately 90\,\%. Now, this matches
well with existing experimental results. This means that the reception
performance is very good in this center region \emph{independent }of
the SIR, i.e., no power control is required and perfect time synchronization
is unnecessary for successful reception.

\subsubsection{Effect of Several Interferers}

\label{sec:N-Interferers}
\begin{figure}
\centering \includegraphics{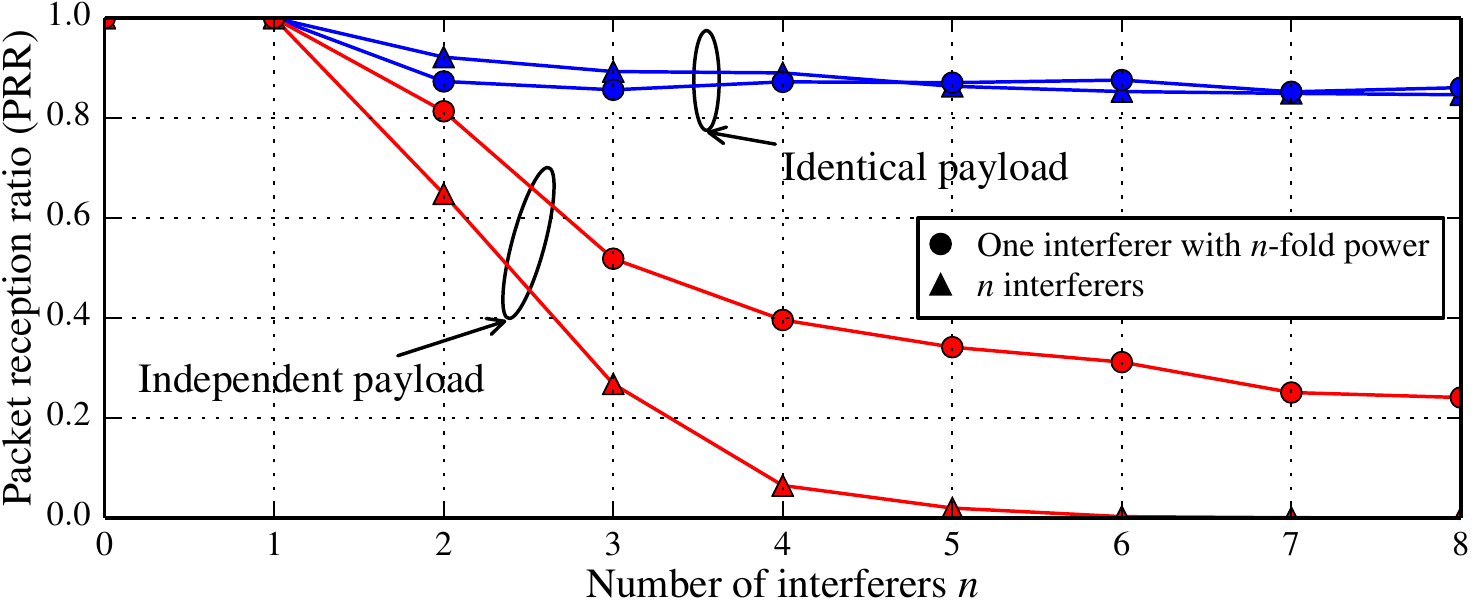}

\caption{\label{fig:nInterferers}Reception ratio for SDD under one strong
interferer or $n$ weaker interferers, but all with equal received
power. For identical payload the difference is small, for independent
payload several interferers are more destructive than one.}
\end{figure}
In this subsection, we consider the effect of one strong interferer
compared to several interferers with the same power when combined,
but evenly distributed across the interferers. We consider the following
scenario: all interferers are time-synchronized ($\tau_{i}=0$), but
each has an i.i.d.~uniform random phase offset $\varphi_{c,i}$ (and
independent payload bits $\beta_{k,i}$ if different content is assumed).
The interference power varies with $\frac{n}{2}P_{\mathrm{SoI}}$
for a number of interferers $n\in\left\{ 1,\ldots,8\right\} $, with
each interferer having a signal power at the receiver of $\frac{1}{2}P_{\mathrm{SoI}}$.

Under the classical capture threshold model both interference types
share the same SIR and thus lead to the same PRR at the receiver.
However, as we observe in \figref{nInterferers}, this is only the
case for identical payload, for independent payload $n$ interferers
prove to be more destructive despite having the same signal power.
While experimental results by Ferrari et al.~suggested this result
\cite[Fig.~12]{Glossy} for identical payload, the root cause is now
explained by our model: a single interferer is more likely affected
by high attenuation ($\cos\varphi_{c}\approx0$) than $n$ independent
interferers, resulting in a higher likelihood of destructive interference.
However, in case of identical payload, even an effective interferer
is still received correctly in 90\,\% of the cases. The observation
for independent payload reveals another problem of SINR models: relying
on the signal power ratio alone discards the crucial effects of each
interferer's offsets.

\subsection{Reception of Interfering Signals with Independent Payload}

\begin{figure}
\centering \subfloat[\label{fig:Eval-USync2-Uncoded}Uncoded transmissions.]{\includegraphics{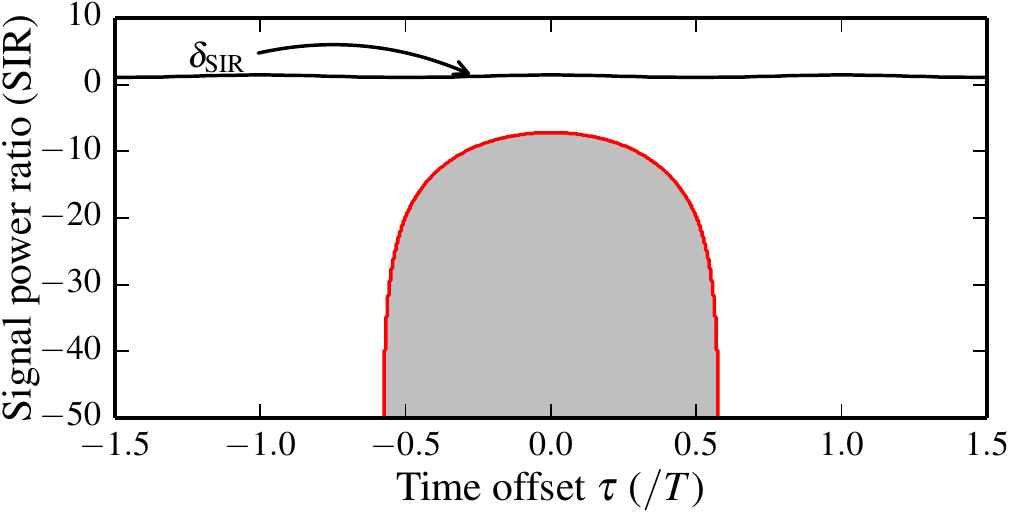}~\includegraphics{figs/ser_contour_AsAu/pdf/cb}

}

\subfloat[\label{fig:Eval-USync2-HDD}DSSS with hard decision decoding (HDD).]{\includegraphics{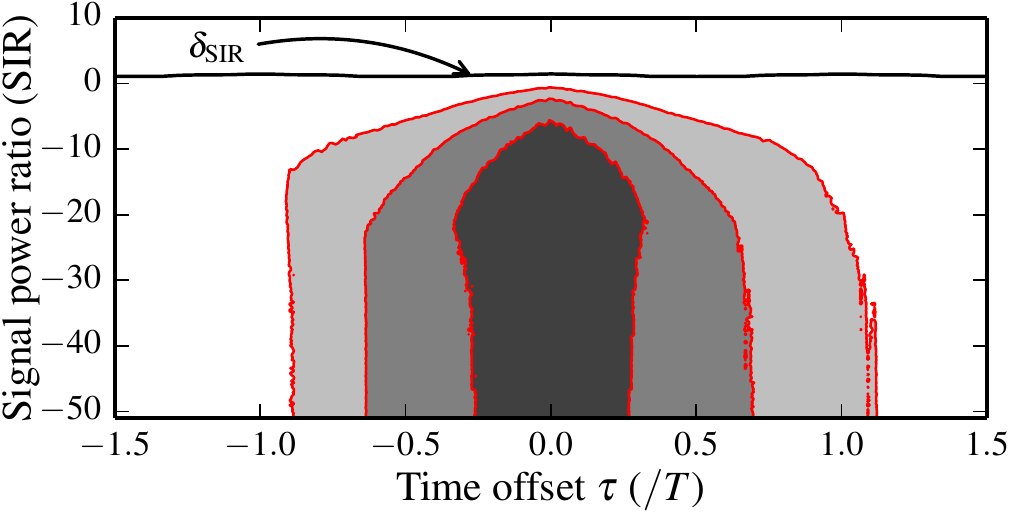}~\includegraphics{figs/ser_contour_AsAu/pdf/cb}

}

\subfloat[\label{fig:Eval-USync-1}DSSS with soft decision decoding (SDD).\label{fig:Eval-USync}]{\includegraphics{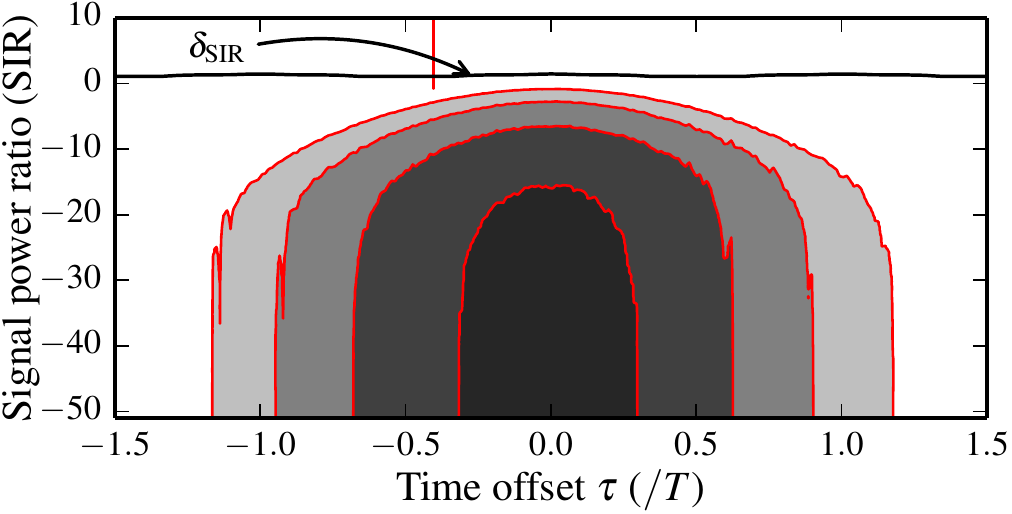}~\includegraphics{figs/ser_contour_AsAu/pdf/cb}

}

\caption{\label{fig:Eval-USync2}Reception regions of an \emph{interfering}
signal with \emph{independent} payload. For reference the reception
threshold for a synchronized signal $\delta_{\mathrm{SIR}}$ (from
\figref{Eval-Indep-Uncoded}) is also shown. (a)~In the uncoded case,
the packet reception ratio in the central region is 20--30\,\% because
phase and time offsets lead to bit flips in the detected packets.
(b)~For HDD, the PRR increases to 60--70\,\% because the coding
helps to mitigate bit errors. (c)~For SDD, the central region enables
a PRR of 80--90\,\% of the interferer's packets.}
\end{figure}
\begin{figure*}
\centering \subfloat[\label{fig:Eye-Uncoded}Bit error rate for uncoded transmissions.]{\includegraphics{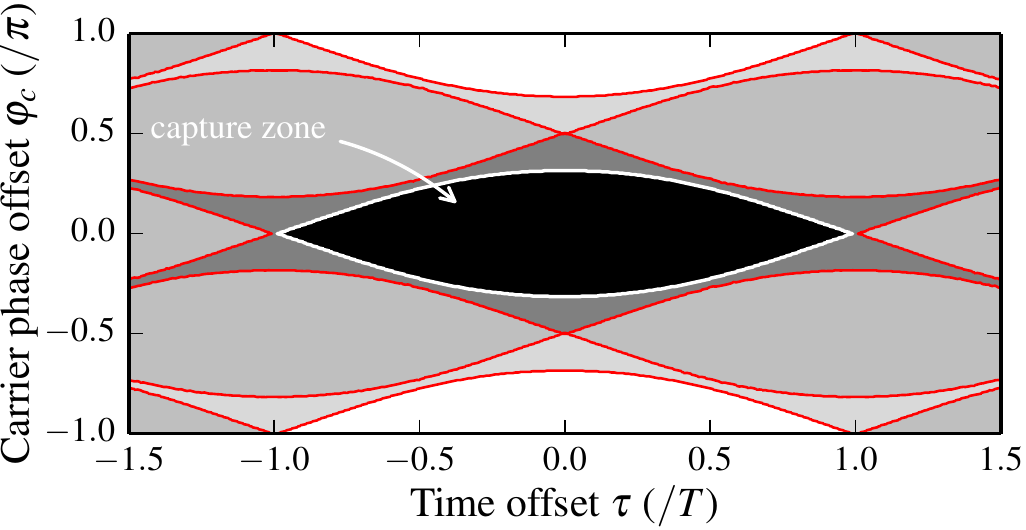}~\includegraphics{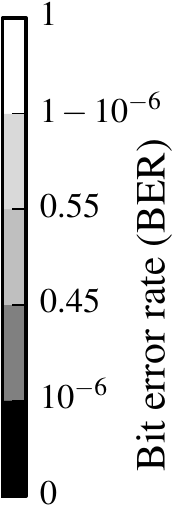}

}

\subfloat[\label{fig:Eye-HDD}Symbol error rate for DSSS/HDD.]{\includegraphics{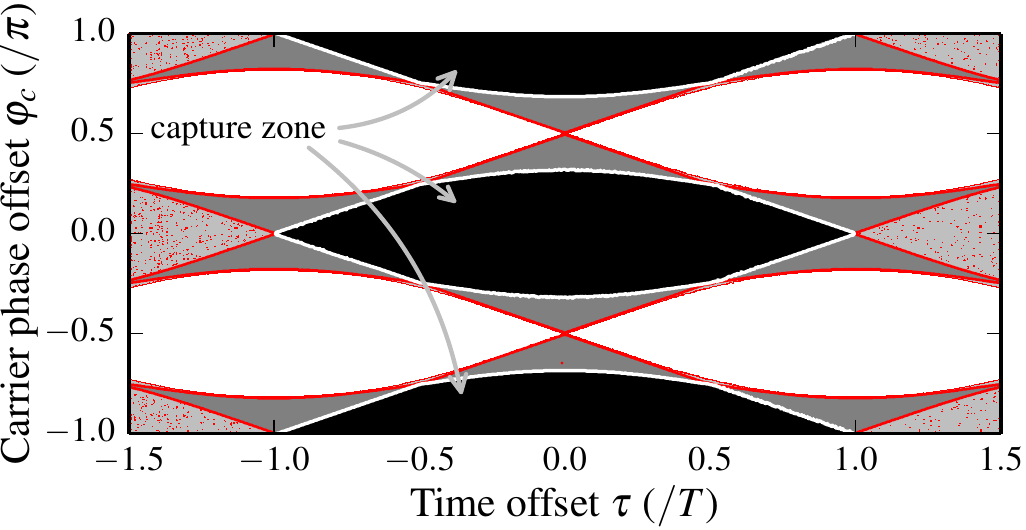}~\includegraphics{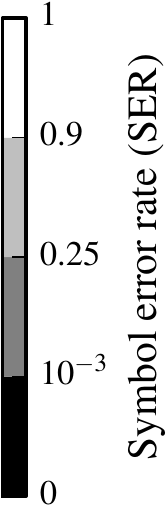}

}

\subfloat[\label{fig:Eye-SDD}Symbol error rate for DSSS/SDD.]{\includegraphics{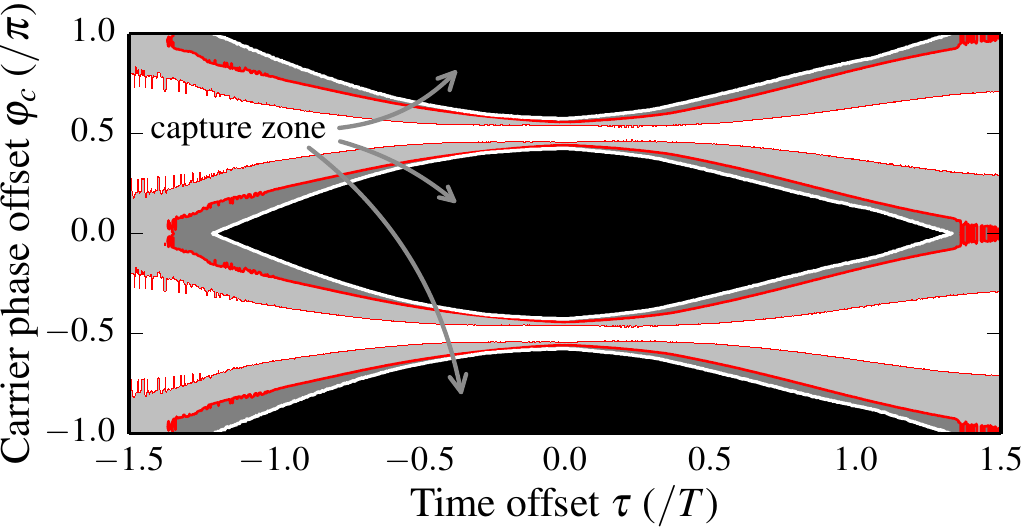}~\includegraphics{figs/1col/figs/ser_contour/pdf/cb}

}

\caption{\label{fig:Eye-Dias}Relation between error rates and signal parameters,
time offset $\tau$ and carrier phase offset $\varphi_{c}$ (with
$\mathrm{SIR}=-40$\,dB). An unsynchronized packet is successfully
received if the parameter combinations fall inside the dark \emph{capture
zones}. (a) For uncoded transmissions, the error rate increases for
phase offsets $\left|\varphi_{c}\right|>\frac{\pi}{4}$. (b) For coded
transmissions and a HDD receiver, the shape of the capture zone is
similar to the case in (a), but a second zone around $\varphi_{c}=\pi$
is present. (c) For coded transmission and SDD, the eye shape is widened,
and an increasing number of parameter combinations result in error-free
transmissions.}
\end{figure*}
Our results explain why and when collision-aware protocols work even
without power control: coding enables the reception of interfering
signals despite carrier phase and time offsets. In this section, we
revisit the case of independent payload but focus our interest now
on the reception of the \emph{interfering} signal, i.e., we treat
the interfering signal $u\left(t\right)$ as the SoI and observe the
reception of $\beta_{k}$ instead of $\alpha_{k}$. Related work by
Pöpper et al.~\cite{11:Wireless-Manip:ESO} shows that for uncoded
systems the reception of interfering signals is indeterministic; in
contrast, we show analytically and experimentally (\secref{Experiments})
that real systems can receive unsynchronized, interfering packets
reliably when using \emph{coded} messages.

\textbf{Uncoded transmissions.} This case is shown in \figref{Eval-USync2-Uncoded}.
In this case, a reception is only successful if bits are not flipped
by either $\varphi_{c}$ or $\varphi_{p}$, and we observe a PRR of
20--30\,\% in the center region ($\mathrm{SIR}<-10$\,dB and $\left|\tau\right|<0.5T$)
in our evaluation. The reason for the poor reception performance is
visible in \figref{Eye-Uncoded}; the acceptable parameter values
of $\tau$ and $\varphi_{c}$ that lead to an error-free packet reception
have tight constraints. The interfering signal must hit into a \emph{capture
zone} defined by the signal parameters, which permits the signal to
have only small time and carrier phase offsets.

\textbf{Hard decision decoding.} In this setting, the PRR in the central
area increases to approx.~60\,\% (\figref{Eval-USync2-HDD}). In
\figref{Eye-HDD}, we see the reason for the increase: while the general
shape is the same, we see a second capture zone around $\varphi_{c}=\pm\pi$.
There are two explanations for this. First, we use the same sliced
bits from the uncoded case as input for DSSS correlation, which thus
possess the same error characteristics. Second, because of the use
of absolute correlation values in the correlation (Eq.~(\ref{eq:DSSSCorr})),
the adverse effect of large phase offsets can be repaired. Specifically,
this means that even if all bits are flipped, the correlation value
is still maximal for the correct chipping sequence. This use of DSSS
thus doubles the PRR of an interfering signal.

\textbf{Soft decision decoding.} Finally, in \figref{Eval-USync},
we see a central area below $\mathrm{SIR}=-23$\,dB and a width of
$0.25T$ that has a PRR for the interfering signal of approx.~90\,\%.
This means that, if the power difference is large enough, a receiver
can ignore a synchronized signal and recover the interfering one despite
its offsets. \figref{Eye-SDD} shows this in terms of the capture
zone. The eye-shaped regions are much wider compared to the other
receiver designs, and especially for the central region with minor
deviations of $\tau$, the SER is negligible. Problems in the reception
only occur for carrier phase offsets such that $\cos\varphi_{c}\approx0$.
These results show that interfering signals can indeed be received,
which helps in collision-aware protocols or other intentional collisions,
e.g., in message manipulation attacks on the physical layer. To validate
this new result, we present an experimental study of such receptions
with real receiver implementations next. 

\section{Experiments}

\label{sec:Experiments}In this section, we provide experimental evidence
that our model accurately captures the behavior of existing receiver
implementations. Since many results in the previous section comply
with existing experimental results (see also \secref{Discussion}),
we focus our efforts on the reception of interfering signals because
this topic is not well covered experimentally in the literature. We
note that we also validated our analytical results with a simulation
model based on the numerical integration of time-discrete signals,
which confirmed the correctness of our model at the symbol and chip
levels. The purpose of this section is to show that our simplifying
assumptions, especially for the receiver model, are justified.

\subsection{Experimental Setup}

To perform this experiment, the requirements for the interferer differ
from the scope of operation of Commercial Off-The-Shelf (COTS) devices.
We need to \emph{(i)} transmit arbitrary symbols on the physical layer,
without restrictions like PHY headers, \emph{(ii)} synchronize to
ongoing transmissions with high accuracy, and \emph{(iii)} schedule
transmissions at a fine time granularity. To meet these requirements,
we implemented a custom software-defined radio based experimental
system.

\subsubsection{Interferer implementation}

To this end, we modified our USRP2-based experimental system RFReact
\cite{WMSL11-3} to recover the timing of the other signal and send
arbitrary IEEE 802.15.4 symbols at controlled time offsets. Because
of its implementation in the USRP2's FPGA, the system is able to tune
the start of transmission with a granularity of 10\,ns and send arbitrary
waveforms. A detailed description of the system can be found in a
technical report \cite{WMSL13}.

\subsubsection{Experimental methodology}

In our experiments, we consider three parties in the network: a standard-compliant
receiver (we monitor the behavior of two implementations to test for
hardware dependencies, Atmel AT86RF230 and TI CC2420), a synchronized
sender $\mathcal{S}$ (a COTS RZ Raven USB), and the interferer $\mathcal{I}$
described above. The procedure is as follows: $\mathcal{S}$ sends
a packet with PHY headers, MAC header, and 8\,byte payload. $\mathcal{I}$
time-synchronizes with this signal and schedules the transmission
of 8 different bytes at the beginning of the payload of $\mathcal{S}$.
The receiver first synchronizes on $\mathcal{S}$ and receives its
header, but experiences a collision in the payload bits. We note that
the receivers do not attempt to correct bit errors, retransmissions
are used for error recovery during normal operation. Damaged packets
are simply detected using the checksum at the end and discarded in
case of failure. For the experiments we reconfigured the devices so
that all packets are recorded, even if the checksums did not match.

We chose values of $\tau$ in $\left(-1.5T;1.5T\right)$ or $\pm750$\,ns
in steps of $10$\,ns; for each time offset $\tau$, we sent 1,000
packets and analyzed the payload detected by the receiver. We derived
the value of $\tau$ empirically, i.e., we chose the point with maximum
PRR in the center as $\tau=0$. We adjusted the transmit power of
$\mathcal{I}$ to result in a SIR of $-40$\,dB to be in the region
of interest.

\subsection{Experimental Results}

\begin{figure}
\centering \subfloat[\label{fig:Exp-Results-PRR}Comparison of packet reception ratios.]{\includegraphics{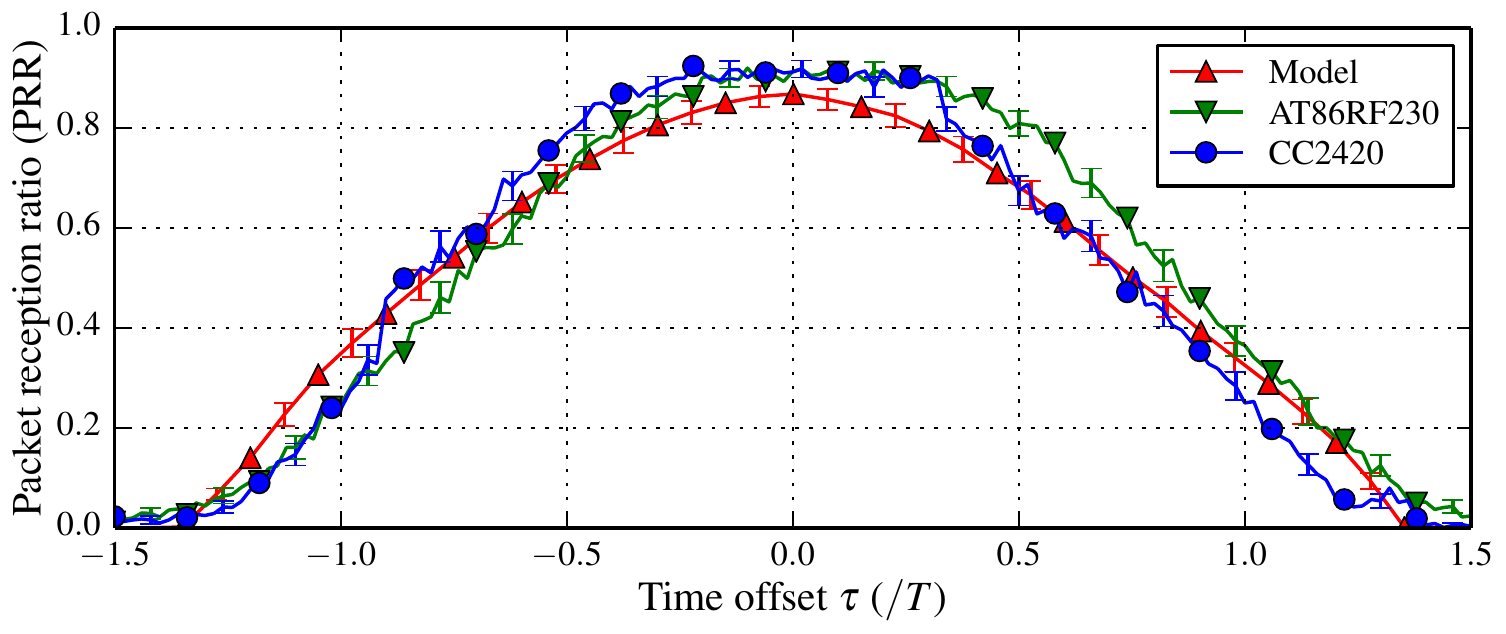}

}

\subfloat[\label{fig:Exp-Res-PRR-std}Comparison of PRR standard deviations.]{\includegraphics{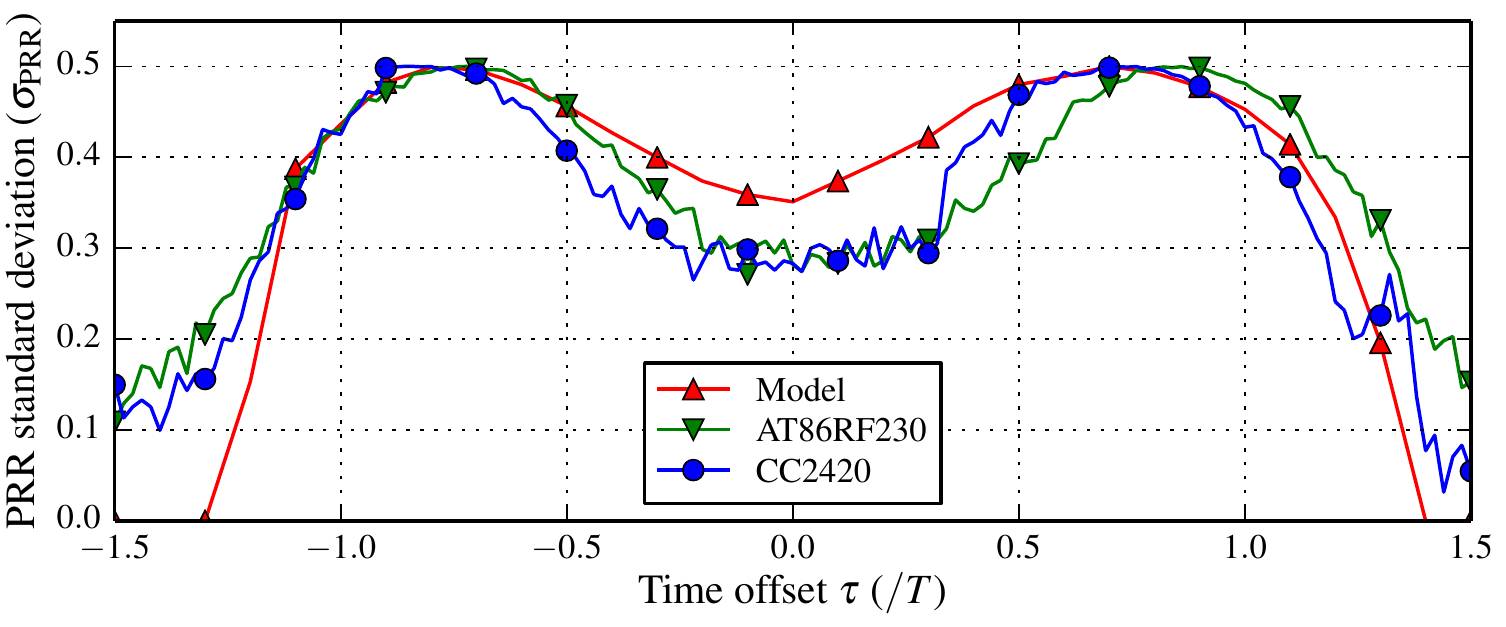}

}

\caption{Experimental results for two receivers in terms of packet reception
(PRR) performance and PRR standard deviation compared to our model.
Both receivers display a behavior that is well-described by the model.}
\end{figure}

We analyze our measurements using two metrics, packet reception ratio
and symbol error rate.

\subsubsection{Packet reception ratio (PRR)}

Based on the received packet data from the experiments, we derive
the PRR as the number of packets with correct payload (of the interferer)
divided by the total number of packets. In other words, we measure
the empirical success probability for a message manipulation attack.
The experimental results for the mean PRR of the two receivers are
shown in \figref{Exp-Results-PRR}. We observe a good fit with the
predictions of our model to both receivers, Atmel AT86RF230 and TI
CC2420. In the central region, the receivers show a slightly better
ability to receive the interfering signal than predicted by our analytical
model. The reason is that our model makes the assumption that no frequency
offset is present and that the receiver does not try to resynchronize
with a stronger signal. However, receivers must be able to tolerate
frequency offsets of up to 100\,kHz \cite[$\S$6.9.4]{ieee802.15.4}
and thus track and possibly correct the phase during the packet reception
process. Yet, as the results show, our assumptions still yield a good
approximation of the real receiver behavior. 

To further validate our model, we perform an analysis of the standard
deviation of the measured PRR values (\figref{Exp-Res-PRR-std}).
In general, the second order statistics follow the non-trivial shape
well. On closer inspection, we observe three regions in the graph.
For $\left|\tau\right|<0.5$, our model slightly overestimates the
standard deviation; the reason is that the PRR performance of the
COTS receivers is better than our model, leading to less variance.
For $0.5<\left|\tau\right|<1.1$, the curves are close to each other.
Finally, in the zone with $\left|\tau\right|>1.1$, the model slightly
underestimates the standard deviation, again because the real receivers
perform better than the model predicts. Still, the model provides
a good approximation of the behavior of widely used receivers for
interfering signals under the assumption of random carrier phase offsets.

\subsubsection{Symbol error rate (SER)}

\begin{figure}
\centering \subfloat[\label{fig:Exp-Res-SER}Comparison of symbol error rates.]{\includegraphics{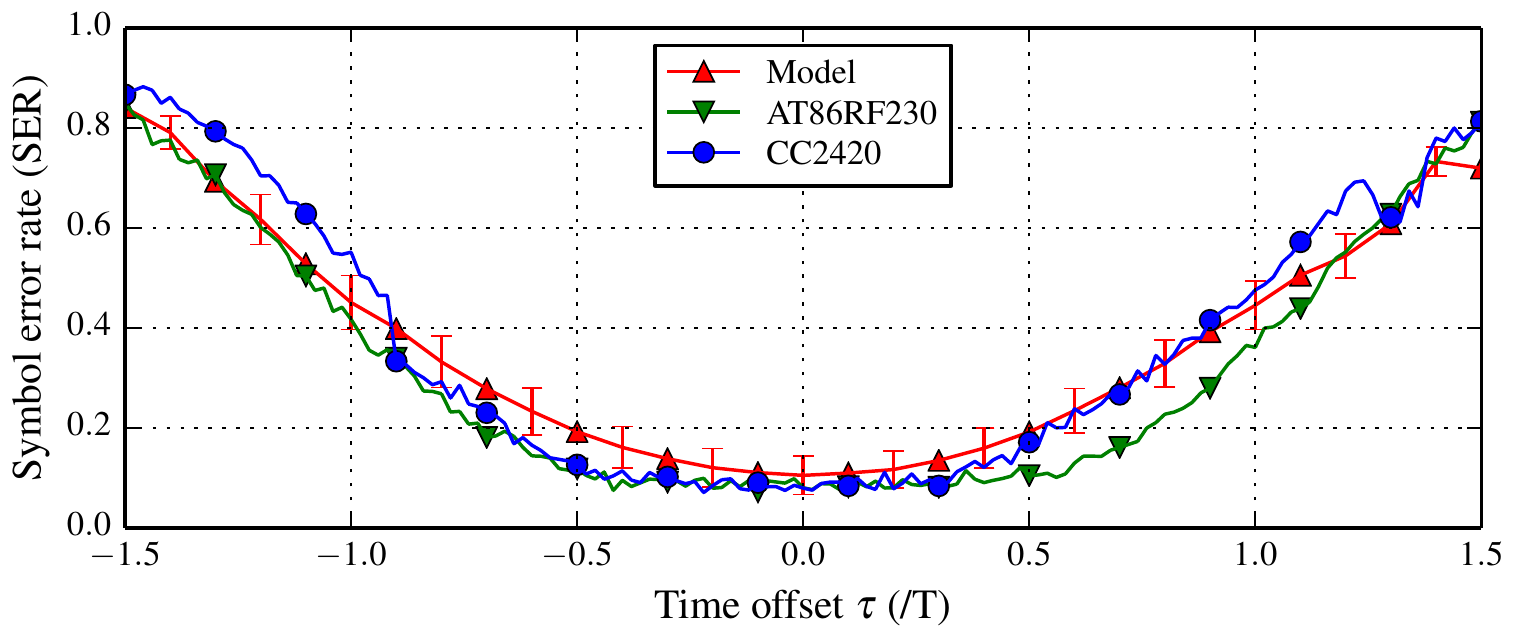}

}

\subfloat[\label{fig:Exp-Res-SER-std}Comparison of SER standard deviations.]{\includegraphics{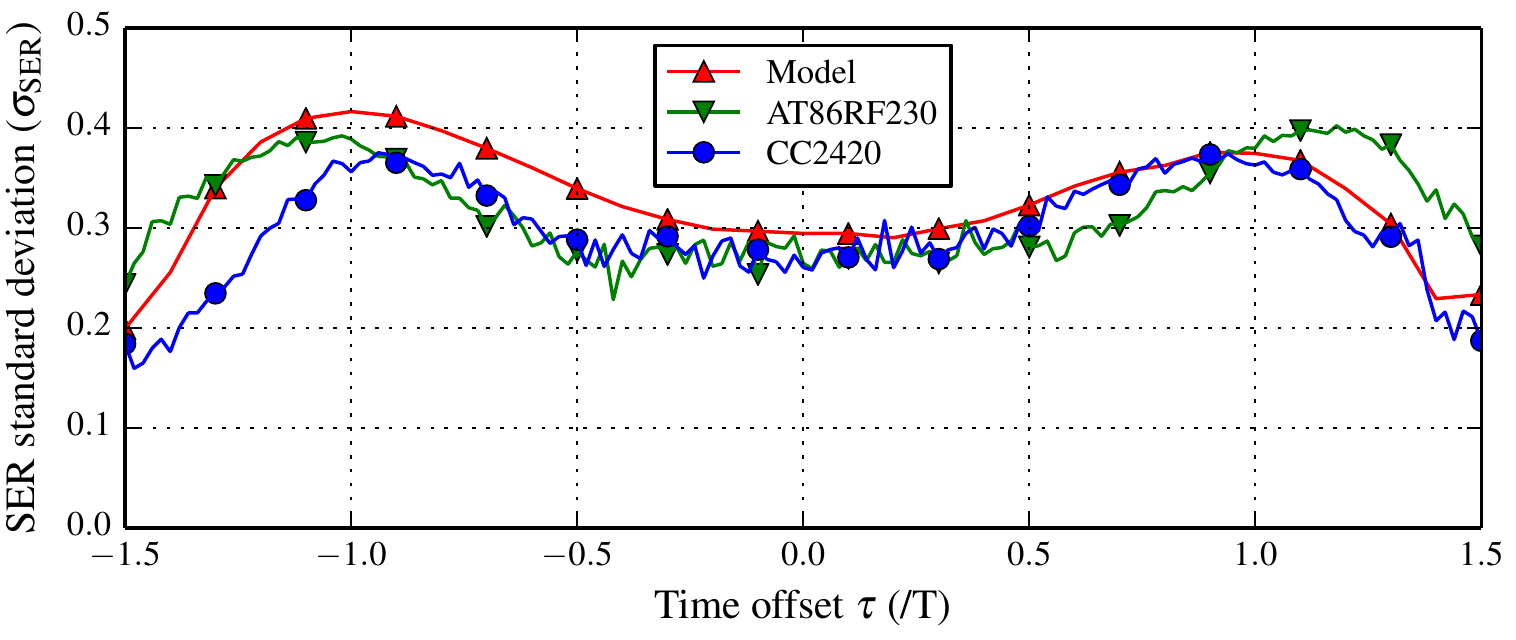}

}

\caption{Comparison of experimentally measured symbol error rates and standard
deviations, and the SER values predicted by our model.}
\end{figure}

We derive the SER by summation of the number of symbol errors across
the payload of all received packets for a given time offset $\tau$,
and divide this sum by the total number of payload symbols. This metric
gives better insights into the causes for packet errors, and provides
another validation for the capture zone. In \figref{Exp-Res-SER},
we observe that the fit is good for the symbol error rate as well,
with a slightly better SER performance for the COTS receivers as expected.
Considering the SER standard deviation (\figref{Exp-Res-SER-std}),
we observe a similar behavior as in the PRR case, the predictions
of the model and the measured results provide a good fit in both,
curve shape and absolute values.

\section{Guidelines for Optimal Parameter Settings}

\label{sec:Discussion}Here, we provide a summary of our main findings
and highlight key conclusions for the design of protocols that leverage
concurrent transmissions. In particular, we summarize how the notion
of capture zones enables engineers and protocol designers to choose
optimal parameter ranges for signal power ratio, time offset, and
carrier phase offset to ensure a successful reception despite collisions.

\subsection{Signal to Interference Ratio SIR}

Our model confirms that when the SoI is above the SIR threshold $\delta_{\mathrm{SIR}}\approx2$\,dB,
then a successful reception is guaranteed (the capture effect). This
is consistent with existing results; for the CC2420 transceiver, Gezer
et al.~\cite{10:154CaptureModel:ISWPC}, Maheshwari et al.~\cite{InfModel},
Dutta et al.~\cite{10:AMAC:SenSys}, and Son et al.~\cite{ExpConcur}
report an experimentally observed threshold of about 3\,dB. Considering
that their channels were not noise-free and that SINR measurements
were collected by the radio transceivers themselves, rather than calibrated
measurement equipment such that inaccuracies may arise, this is consistent
with our results. If it can be ensured that the stronger signal arrives
first and the synchronization process succeeds, the SIR-based capture
threshold is a valid model for receiver behavior.

A different matter is the case when the SoI is located in the negative
SIR regime, i.e., the interfering signal is stronger than the SoI.
This situation occurs if an interferer is closer to the receiver or
the synchronization process fails because of a collision during the
preamble (which is the case, for example, for the collision-aware
flooding protocols). Our model gives better insights in this situation
and shows that a reception may still be possible \emph{no matter what
SIR}, given that the interfering signal parameters are in the capture
zone as defined by the time offset $\tau$ and carrier phase offset
$\varphi_{c}$. Valid settings for these parameters are discussed
below.

\subsection{Time Offset $\tau$}

As a guideline derived from the capture zone, the time offset $\tau$
should be below $T/2$ for successful concurrent transmissions with
identical content, which translates to $250$\,ns for IEEE 802.15.4.
Thus network flooding protocols, for example Glossy, should aim to
keep the transmission start time error below this value to ensure
a desired PRR above 75\,\%. If $\tau<200$\,ns can be ensured, the
achievable PRR is approximately 90\,\%. We note that this ensures
worst-case performance (i.e., the SoI is always in the negative SIR
regime). The actual performance may be higher in situations with positive
SIR or successful synchronization.

\subsection{Carrier Phase Offset $\varphi_{c}$}

If the carrier frequency offset at the receiver can be precisely controlled
by the senders, there are several options. Interferers can choose
$\varphi_{c}\approx\pm\frac{\pi}{2}$ to minimize their effect on
the SoI, reducing their influence to signal demodulation. On the other
hand, interferers could aim for the capture zone (e.g., $\left|\varphi_{c}\right|<0.4\pi$
or $\left|\varphi_{c}\right|>0.6\pi$ for $\tau=0$ and SDD) to ensure
that their signal is received without errors. There are, however,
few approaches in the literature that aim to exploit this. The reason
is that the carrier phase at another physical location is hard to
predict except in static and free space scenarios because of fading
and multipath effects. Pöpper et al.~\cite{11:Wireless-Manip:ESO}
show for uncoded QPSK that carrier phase offsets are the major hindrance
for a (malicious) interferer to control the bit decisions. In contrast,
the results based on our model suggest that such precise phase control
is not necessary when DSSS is used, and that intentional message manipulations
by deliberate interference are indeed a real threat \cite{MMA802.15.4:MMB:2012}.

\subsection{Number of Concurrent Interferers}

Our results in \secref{N-Interferers} explain why the number of interferers
only has a small impact on reception performance for concurrent transmissions
using identical payload. Ferrari et al.~\cite{Glossy} observed this
behavior in their experiments, achieving a stable PRR above 98\,\%
for 2--10 concurrent transmissions. Maheshwari et al.~\cite{InfModel}
observed that the SIR threshold is not varying with an increasing
number of interferers. On the other hand, Lu and Whitehouse~\cite{FlashFlood}
reported a decreasing PRR when the number of interferers is increased.
However, the Flash Flooding protocol relies on capture, such that
increased time offsets may also influence the results. Some related
work claims that a greater number of concurrent transmitters cause
problems (Wang et al.~\cite{13:ScaleFlood:ToN}, Doddavenkatappa
et al.~\cite{13:Splash:NSDI}) because ``the probability of the
maximum time displacement across different transmitters exceeding
the required threshold for constructive interference'' may increase.
Our model shows that these protocol-related issues should be addressed
with more precise timing synchronization across the network. For independent
payload, we show that 2--3 interferers are sufficient to reduce the
PRR significantly. This confirms the effect reported by Gezer et al.~\cite{10:154CaptureModel:ISWPC}
that the PRR decreases with an increasing number of interferers.

\section{Conclusion}

In this article, we presented the first comprehensive analytical model
for concurrent transmissions over a wireless channel. As shown in
an extensive parameter space exploration, the model recovers insights
from experimental results found in the literature and going beyond
that, explains the root causes for successful concurrent transmissions
exploited in a new generation of sensor network protocols that intentionally
generate collisions to increase network throughput or to reduce latency.
Our results reveal that power capture alone is not sufficient to explain
the performance of such protocols. Rather, coding is an essential
factor in the success of these protocols because it crucially widens
the capture zone of acceptable signal offsets, increasing the probability
of successful reception. Finally, our experimental study of packet
reception under collisions shows a good fit and reinforces the validity
of our model; as a further contribution, we demonstrated the feasibility
of message manipulation attacks over the air experimentally.
\bibliographystyle{ieeetr}
\bibliography{refs}

\begin{thebibliography}{10}

\bibitem{CaptureFM}
K.~Leentvaar and J.~Flint, ``The capture effect in {FM} receivers,'' {\em IEEE
  Trans.~Commun.}, vol.~24, pp.~531--539, May 1976.

\bibitem{08:MultiCapture:JSAC}
J.~Foo and D.~Huang, ``Multiuser diversity with capture for wireless networks:
  Protocol and performance analysis,'' {\em IEEE J.~Sel.~Areas Commun.},
  vol.~26, pp.~1386--1396, Oct. 2008.

\bibitem{UnderstandRFInterf}
R.~Gummadi, D.~Wetherall, B.~Greenstein, and S.~Seshan, ``Understanding and
  mitigating the impact of {RF} interference on 802.11 networks,'' in {\em
  Proc.~ACM SIGCOMM~'07}, pp.~385--396, Sept. 2007.

\bibitem{04:WLANCapture:ICNP}
A.~Kochut, A.~Vasan, A.~Shankar, and A.~Agrawala, ``Sniffing out the correct
  physical layer capture model in 802.11b,'' in {\em Proc.~IEEE ICNP~'04},
  pp.~252--261, Oct. 2004.

\bibitem{Capture11a}
J.~Lee, W.~Kim, S.-J. Lee, D.~Jo, J.~Ryu, T.~Kwon, and Y.~Choi, ``An
  experimental study on the capture effect in 802.11a networks,'' in {\em
  Proc.~ACM WinTECH~'07}, pp.~19--26, Sept. 2007.

\bibitem{10:154CaptureModel:ISWPC}
C.~Gezer, C.~Buratti, and R.~Verdone, ``Capture effect in {IEEE} 802.15.4
  networks: Modelling and experimentation,'' in {\em Proc.~IEEE ISWPC~'10},
  pp.~204--209, May 2010.

\bibitem{InfModel}
R.~Maheshwari, S.~Jain, and S.~R. Das, ``A measurement study of interference
  modeling and scheduling in low-power wireless networks,'' in {\em Proc.~ACM
  SenSys~'08}, pp.~141--154, Nov. 2008.

\bibitem{ExpConcur}
D.~Son, B.~Krishnamachari, and J.~Heidemann, ``Experimental study of concurrent
  transmission in wireless sensor networks,'' in {\em Proc.~ACM SenSys~'06},
  pp.~237--250, Nov. 2006.

\bibitem{CMAC}
M.~Sha, G.~Xing, G.~Zhou, S.~Liu, and X.~Wang, ``{C-MAC}: Model-driven
  concurrent medium access control for wireless sensor networks,'' in {\em
  Proc.~IEEE INFOCOM~'09}, pp.~1845--1853, Apr. 2009.

\bibitem{08:HarnessExposed:NSDI}
M.~Vutukuru, K.~Jamieson, and H.~Balakrishnan, ``Harnessing exposed terminals
  in wireless networks,'' in {\em Proc.~USENIX NSDI~'08}, pp.~59--72, Apr.
  2008.

\bibitem{10:AMAC:SenSys}
P.~Dutta, S.~Dawson-Haggerty, Y.~Chen, C.-J.~M. Liang, and A.~Terzis, ``Design
  and evaluation of a versatile and efficient receiver-initiated link layer for
  low-power wireless,'' in {\em Proc.~ACM SenSys~'10}, pp.~1--14, ACM, Nov.
  2010.

\bibitem{ACKCollisions}
P.~Dutta, R.~Musaloiu-E., I.~Stoica, and A.~Terzis, ``Wireless {ACK} collisions
  not considered harmful,'' in {\em Proc.~ACM SIGCOMM HotNets-VII}, pp.~4:1--6,
  Oct. 2008.

\bibitem{14:CountCC:IPSN}
D.~Wu, C.~Dong, S.~Tang, H.~Dai, and G.~Chen, ``Fast and fine-grained counting
  and identification via constructive interference in {WSNs},'' in {\em
  Proc.~IEEE\slash{}ACM IPSN '14}, pp.~191--202, Apr. 2014.

\bibitem{13:Splash:NSDI}
M.~Doddavenkatappa, M.~C. Chan, and B.~Leong, ``Splash: Fast data dissemination
  with constructive interference in wireless sensor networks,'' in {\em
  Proc.~USENIX NSDI~'13}, pp.~269--282, Apr. 2013.

\bibitem{Glossy}
F.~Ferrari, M.~Zimmerling, L.~Thiele, and O.~Saukh, ``Efficient network
  flooding and time synchronization with {Glossy},'' in {\em Proc.~IPSN~'11},
  pp.~73--84, Apr. 2011.

\bibitem{FlashFlood}
J.~Lu and K.~Whitehouse, ``{Flash Flooding}: Exploiting the capture effect for
  rapid flooding in wireless sensor networks,'' in {\em Proc.~IEEE
  INFOCOM~'09}, pp.~2491--2499, Apr. 2009.

\bibitem{13:ScaleFlood:ToN}
Y.~Wang, Y.~He, X.~Mao, Y.~Liu, and X.~Li, ``Exploiting constructive
  interference for scalable flooding in wireless networks,'' {\em
  IEEE\slash{}ACM Trans.~Netw.}, vol.~21, pp.~1880--1889, Dec. 2013.

\bibitem{Triggercast}
Y.~Wang, Y.~Liu, Y.~He, X.-Y. Li, and D.~Cheng, ``Disco: Improving packet
  delivery via deliberate synchronized contructive interference,'' {\em IEEE
  Trans.~Parallel Distrib.~Syst.}, 2014.
\newblock To appear.

\bibitem{CaptureAdHoc}
N.~Santhapuri, S.~Nelakuditi, and R.~R. Choudhury, ``On spatial reuse and
  capture in ad hoc networks,'' in {\em Proc.~IEEE WCNC~'08}, pp.~1628--1633,
  Apr. 2008.

\bibitem{DelayCap}
D.~Davis and S.~Gronemeyer, ``Performance of slotted {ALOHA} random access with
  delay capture and randomized time of arrival,'' {\em IEEE Trans.~Commun.},
  vol.~28, pp.~703--710, May 1980.

\bibitem{11:Wireless-Manip:ESO}
C.~P\"{o}pper, N.~O. Tippenhauer, B.~Danev, and S.~\v{C}apkun, ``Investigation
  of signal and message manipulations on the wireless channel,'' in {\em
  Computer Security -- ESORICS 2011}, no.~6879 in LNCS, pp.~40--59, Springer,
  Sept. 2011.

\bibitem{05:Capture:EmNetS}
K.~Whitehouse, A.~Woo, F.~Jiang, J.~Polastre, and D.~Culler, ``Exploiting the
  capture effect for collision detection and recovery,'' in {\em Proc.~IEEE
  EmNetS-II}, pp.~45--52, May 2005.

\bibitem{TalkTogether}
D.~Yuan and M.~Hollick, ``Let's talk together: Understanding concurrent
  transmissions in wireless sensor networks,'' in {\em Proc.~IEEE LCN~'13},
  pp.~219--227, Oct. 2013.

\bibitem{ModelGlossy}
M.~Zimmerling, F.~Ferrari, L.~Mottola, and L.~Thiele, ``On modeling low-power
  wireless protocols based on synchronous packet transmissions,'' in {\em
  Proc.~IEEE MASCOTS~'13}, pp.~546--555, Aug. 2013.

\bibitem{PPR}
K.~Jamieson and H.~Balakrishnan, ``{PPR}: Partial packet recovery for wireless
  networks,'' in {\em Proc.~ACM SIGCOMM~'07}, pp.~409--420, Sept. 2007.

\bibitem{BitErrDistLPWN}
F.~Schmidt, M.~Ceriotti, and K.~Wehrle, ``Bit error distribution and mutation
  patterns of corrupted packets in low-power wireless networks,'' in {\em
  Proc.~ACM WiNTECH~'13}, pp.~49--56, Sept. 2013.

\bibitem{12:RedCliff:INFC}
K.~Wu, H.~Tan, H.-L. Ngan, Y.~Liu, and L.~M. Ni, ``Chip error pattern analysis
  in {IEEE} 802.15.4,'' {\em IEEE Trans.~Mob.~Comput.}, vol.~11, pp.~543--552,
  Apr. 2012.

\bibitem{GuptaKumar}
P.~Gupta and P.~R. Kumar, ``The capacity of wireless networks,'' {\em IEEE
  Trans.~Inf.~T.}, vol.~46, pp.~388--404, May 2000.

\bibitem{ModelInterf}
P.~Cardieri, ``Modeling interference in wireless ad hoc networks,'' {\em IEEE
  Commun.~Surveys Tutorials}, vol.~12, pp.~551--572, Sept. 2010.

\bibitem{IntfModels}
A.~Iyer, C.~Rosenberg, and A.~Karnik, ``What is the right model for wireless
  channel interference?,'' {\em IEEE Trans.~Wireless Commun.}, vol.~8,
  pp.~2662--2671, June 2009.

\bibitem{Reorder}
J.~Manweiler, N.~Santhapuri, S.~Sen, R.~R. Choudhury, S.~Nelakuditi, and
  K.~Munagala, ``Order matters: Transmission reordering in wireless networks,''
  {\em IEEE/ACM Trans.~Netw.}, vol.~20, pp.~353--366, Apr. 2012.

\bibitem{07:DigiCom:book}
J.~Proakis and M.~Salehi, {\em Digital Communications}.
\newblock New York, NY: McGraw-Hill, 5th~ed., Nov. 2007.

\bibitem{ieee802.15.4}
``{IEEE Standard 802 Part 15.4}: Wireless medium access control and physical
  layer specifications for low-rate {WPANs}, {Sept.} 2006..''

\bibitem{79:MSK:ComMag}
S.~Pasupathy, ``Minimum shift keying: A spectrally efficient modulation,'' {\em
  IEEE Comm.~Mag.}, vol.~17, pp.~14--22, July 1979.

\bibitem{05:GNUDeEn:TR}
T.~Schmid, ``{GNU Radio} 802.15.4 en- and decoding,'' Tech. Rep.
  TR-UCLA-NESL-200609-06, UCLA NESL, 2005.

\bibitem{JainPerf}
R.~K. Jain, {\em The Art of Computer Systems Performance Analysis: Techniques
  for Experimental Design, Measurement, Simulation, and Modeling}.
\newblock Hoboken, NJ: John Wiley \& Sons, Apr. 1991.

\bibitem{WC:Rappaport}
T.~S. Rappaport, {\em Wireless Communications: Principles and Practice}.
\newblock Upper Saddle River, NJ: Prentice-Hall, 2nd~ed., Apr. 1996.

\bibitem{PoiselJamming}
R.~A. Poisel, {\em Modern Communications Jamming: Principles and Techniques}.
\newblock Boston, MA: Artech House Publishers, Nov. 2003.

\bibitem{WMSL11-3}
M.~Wilhelm, I.~Martinovic, J.~B. Schmitt, and V.~Lenders, ``{WiSec}\,'11 demo:
  {RFReact}---a real-time capable and channel-aware jamming platform,'' {\em
  {SIGMOBILE} Mobile Comp.~Commun.~Rev.}, vol.~15, pp.~41--42, Nov. 2011.

\bibitem{WMSL13}
M.~Wilhelm, I.~Martinovic, J.~B. Schmitt, and V.~Lenders, ``Air dominance in
  sensor networks: Guarding sensor motes using selective interference,'' Tech.
  Rep. arXiv:1305.4038, TU~Kaiserslautern, Germany, May 2013.

\bibitem{MMA802.15.4:MMB:2012}
M.~Wilhelm, J.~B. Schmitt, and V.~Lenders, ``Practical message manipulation
  attacks in {IEEE} 802.15.4 wireless networks,'' in {\em Workshop Proc.~MMB
  '12}, pp.~29--31, Mar. 2012.

\end{thebibliography}

\global\long\def\Sif#1{S_{k}^{I}\left(#1\right)}
\global\long\def\Sqf#1{S_{k}^{Q}\left(#1\right)}
\global\long\def\IntI{\int_{\left(2k-1\right)T}^{\left(2k+1\right)T}}
\global\long\def\f{f\left(t\right)}
\global\long\def\F{F\left(t\right)}
\global\long\def\puq{\underline{\varphi_{p}^{Q}}}

\fleqn %% Appendices only: float left to all derivations to get a nice alignment and more space 

\appendices{}

\section{Integrating Rectangle Pulses}

\label{apx:IntegratingRect}
\begin{figure}
\includegraphics{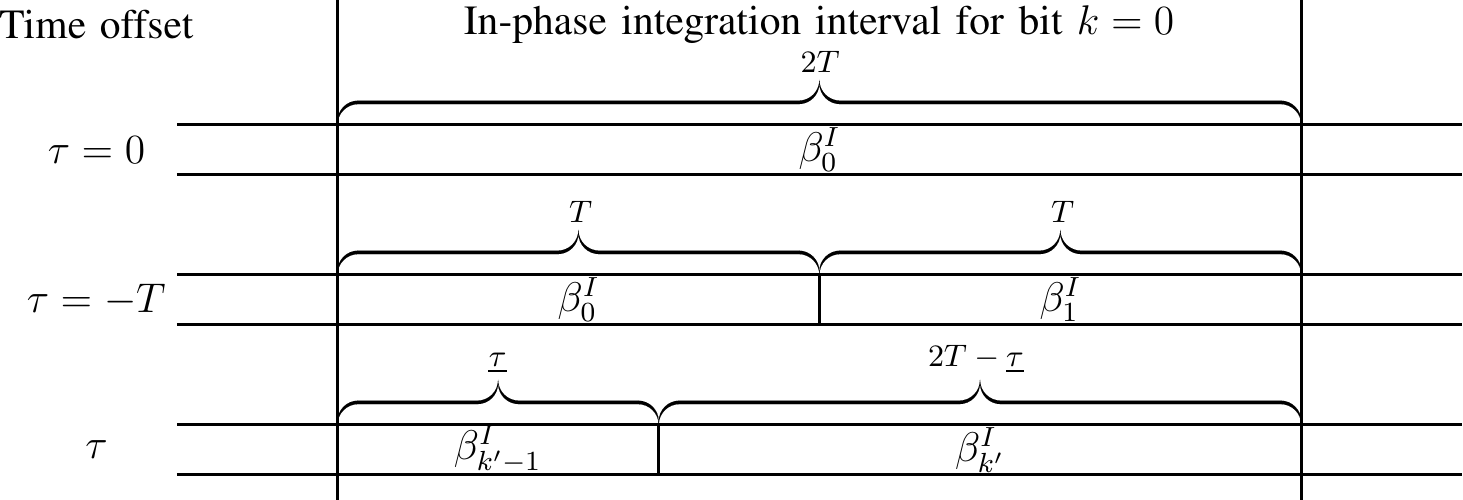}\caption{\label{fig:ExampleRect-Integration}Examples of active bits in the
integration interval for the $I$-bit $k$. For $\tau=0$, the only
active bit in the integration interval is $\beta_{0}^{I}$. When the
signal starts half a bit-length too early ($\tau=-T$), there are
two bits $\beta_{0}^{I}$ and $\beta_{1}^{I}$ that contribute equally
to the bit decision, both are active for a duration of $T$. In the
general case of a time offset $\tau$, there are two active bits with
indices $\beta_{k^{\prime}-1}^{I}$ and $\beta_{k^{\prime}}^{I}$,
with an active time duration of $\protect\tu$ and $2T-\protect\tu$,
respectively.}
\end{figure}
A central equation for deriving the influence of individual bits on
the demodulator output is the integration of the superposition of
time shifted unit pulses $\Pi\left(t\right)$ (defined in \eqref{Pulse}).
This is especially important because of signal time offsets $\tau$
that shift the pulses relative to the integration interval. Situations
that arise are shown in \figref{ExampleRect-Integration}.

To this end, we first derive the general result to the integration
over one bit interval $k$ for arbitrary, integrable functions $f\left(t\right)$.
We consider two variants, the integration of in-phase bits, and the
special case of integrating quadrature-phase bits in the bounds of
$I$-bits (which happens when $Q$-bits leak into the $I$-phase),
i.e., \begin{ceqn} 
\begin{flalign*}
\Sif f & =\IntI b_{I}\left(t-\tau\right)\f dt\\
\Sqf f & =\IntI b_{Q}\left(t-\tau\right)\f dt.
\end{flalign*}
\end{ceqn}Our approach is to split each equation into two parts where
the unit pulse is the constant 1 function to simplify the equations.
Since only one pulse is active at any point in time, such splitting
is possible.

\subsection{Integrating Bit Pulses During the $I$ Integration Interval}

\subsubsection{Integration of $I$-bits}

To perform the integration, we first derive the two indices that have
active pulses during the integration interval. The shift introduced
by $\tau$ lead to the two new bits with indices $k^{\prime}=k-\left\lfloor \frac{\tau}{2T}\right\rfloor $
and $k^{\prime}-1$. The remaining time offset inside the selected
bits is $\tu=\tau-2k_{\tau}T$, i.e., each of the two bits is active
for the time interval $\tu$ and $2T-\tu$, respectively. Because
of this definition, the values of $\tu$ are restricted to the interval
$\left(0,2T\right)$---negative values would activate previous bits,
which is prevented by the floor operation.

For the in-phase component, we derive 
\begin{flalign*}
 & \Sif f\\
 & =\IntI b_{I}\left(t-\tau\right)\f dt\\
 & =\IntI b_{I}\left(t-2k_{\tau}T-\tu\right)\f dt\\
 & =\IntI\sum_{k=-\infty}^{\infty}\bi\Pi\left(\frac{t-\tu-\left(k+k_{\tau}\right)2T}{2T}\right)\f dt\\
 & \intertext{\text{Re-labeling the bit indices \ensuremath{k}~to \ensuremath{k^{\prime}}~(note: positive time shifts lead to negative index shifts)}}\\
 & =\IntI\left(\bkni\Pi\left(\frac{t-\tu-\left(k-1\right)2T}{2T}\right)+\bki\Pi\left(\frac{t-\tu-2kT}{2T}\right)\right)\f dt\\
 & =\bkni\IntI\Pi\left(\frac{t-\tu-\left(k-1\right)2T}{2T}\right)\f dt+\bki\IntI\Pi\left(\frac{t-\tu-2kT}{2T}\right)\f dt\\
 & \intertext{\text{Use the fact that the shifted pulses are zero during parts of the integration interval}}\\
 & =\bkni\int_{\left(2k-1\right)T}^{\left(2k-1\right)T+\tu}\Pi\left(\frac{t-\tu-\left(k-1\right)2T}{2T}\right)\f dt+\bki\int_{\left(2k-1\right)T+\tu}^{\left(2k+1\right)T}\Pi\left(\frac{t-\tu-2kT}{2T}\right)\f dt\\
 & \intertext{\text{The \ensuremath{\Pi}~pulses are constant 1 in the new integration intervals}}\\
 & =\bkni\int_{\left(2k-1\right)T}^{\left(2k-1\right)T+\tu}\f dt+\bki\int_{\left(2k-1\right)T+\tu}^{\left(2k+1\right)T}\f dt\\
 & =\bkni\left[\F\vphantom{\prod}\right]_{2kT-T}^{2kT-T+\tu}+\bki\left[\F\vphantom{\prod}\right]_{2kT-T+\tu}^{2kT+T}
\end{flalign*}
If the function to integrate is the constant 1 function ($f\left(t\right)=1$),
then we derive \begin{ceqn} 
\begin{flalign}
\Sif 1 & =\tu\bkni+\left(2T-\tu\right)\bki\label{eq:Rect1-I}
\end{flalign}
\end{ceqn}

\subsubsection{Integration of $Q$-bits}

\label{apx:Int-Qbits}When $Q$ bits leak into the in-phase, we have
the consider the additional shift of $T$ due to the staggering of
bits in the MSK modulation. We provide the derivation of this special
case here. First, we substitute the timing offset $\tau$ with $\tau^{Q}=\tau+T$
to accommodate of the staggering. Second, the bit indices must be
re-adjusted because of the shift; the new index is denoted by $k^{Q\prime}=k-\left\lfloor \left(\tau+T\right)/2T\right\rfloor $.
For the case of the constant 1 function, we derive then \begin{ceqn}
\begin{flalign}
\Sqf 1 & =\tuq\bknq+\left(2T-\tuq\right)\bkq.\label{eq:Rect1-Q}
\end{flalign}
\end{ceqn}

\subsection{Deriving Special Cases: \textmd{\normalsize{}$\protect\Sif{\cos2\omega_{p}t}$
and }$\protect\Sqf{\cos2\omega_{p}t}$\label{sec:Cos-I-in-Q}}

\subsubsection{Integration of $I$-bits}

We derive the result of bit pulse integration for this special case.
\begin{flalign*}
 & \Sif{\cos2\omega_{p}t}\\
 & =\IntI b_{I}\left(t-\tau\right)\cos2\omega_{p}t\: dt\\
 & =\bkni\left[\frac{1}{2\omega_{p}}\sin2\omega_{p}t\vphantom{\prod}\right]_{\left(2k-1\right)T}^{\left(2k-1\right)T+\tu}+\bki\left[\frac{1}{2\omega_{p}}\sin2\omega_{p}t\vphantom{\prod}\right]_{\left(2k-1\right)T+\tu}^{\left(2k+1\right)T}\\
 & =\frac{\bkni}{2\omega_{p}}\left[\sin2\omega_{p}t\vphantom{\prod}\right]_{\left(2k-1\right)T}^{\left(2k-1\right)T+\tu}+\frac{\bki}{2\omega_{p}}\left[\sin2\omega_{p}t\vphantom{\prod}\right]_{\left(2k-1\right)T+\tu}^{\left(2k+1\right)T}
\end{flalign*}
Performing the integration results in (we denote $\omega_{p}\tu=\underline{\varphi_{p}}$):
\begin{flalign*}
 & =\frac{\bkni}{2\omega_{p}}\left[\sin\left(\left(2k-1\right)\pi+2\underline{\varphi_{p}}\right)-\sin\left(\left(2k-1\right)\pi\right)\right]\\
 & \hphantom{=}+\frac{\bki}{2\omega_{p}}\left[\sin\left(\left(2k+1\right)\pi\right)-\sin\left(\left(2k-1\right)\pi+2\underline{\varphi_{p}}\right)\right]\\
 & =\frac{\bkni}{2\omega_{p}}\left[\sin\left(-\pi+2\underline{\varphi_{p}}\right)-\sin\left(-\pi\right)\right]+\frac{\bki}{2\omega_{p}}\left(\sin\pi+\sin2\underline{\varphi_{p}}\right)\\
 & =-\frac{\bkni}{2\omega_{p}}\sin2\underline{\varphi_{p}}+\frac{\bki}{2\omega_{p}}\sin2\underline{\varphi_{p}}\\
 & =\sin2\underline{\varphi_{p}}\left(-\frac{\bkni}{2\omega_{p}}+\frac{\bki}{2\omega_{p}}\right)\\
 & \intertext{\text{Using \ensuremath{\sin}2\ensuremath{\underline{\varphi_{p}}} = \ensuremath{\sin\left(2\omega_{p}\left(\tau-2k_{\tau}T\right)\right)} = \ensuremath{\sin\left(2\varphi_{p}-2k_{\tau}\pi\right)} = \ensuremath{\sin}2\ensuremath{\varphi_{p}}}}\\
 & =-\frac{1}{2\omega_{p}}\sin2\varphi_{p}\left(\bkni-\bki\right)
\end{flalign*}
The overall result is \begin{ceqn} 
\begin{flalign}
\Sif{\cos2\omega_{p}t} & =-\frac{1}{2\omega_{p}}\sin2\varphi_{p}\left(\bkni-\bki\right)\label{eq:RectCos-I}
\end{flalign}
\end{ceqn}

\subsubsection{Integration of $Q$-bits}

In this case, the use of $\tau^{Q}$ leads to a different phase shift
$\varphi_{p}^{Q}=\omega_{p}\left(\tau+T\right)=\omega_{p}\tau+\frac{\pi T}{2T}=\varphi_{p}+\frac{\pi}{2}$
that leads to changes in the integration. Using the following two
simplifications the derivation can be performed analogously to the
previous subsection.

\begin{flalign*}
\sin2\puq & =\sin2\omega_{p}\tuq=\sin\left(2\frac{\pi}{2T}\left(\tau^{Q}-2k_{\tau}^{Q}T\right)\right)=\sin\left(\frac{\tau^{Q}\pi}{T}-2k_{\tau}^{Q}\pi\right)=\sin2\varphi_{p}^{Q}\\
 & \intertext{\text{and }}\\
\sin2\varphi_{p}^{Q} & =\sin\left(2\varphi_{p}+\pi\right)=-\sin2\varphi_{p}
\end{flalign*}
The overall result is \begin{ceqn} 
\begin{flalign}
\Sqf{\cos2\omega_{p}t} & =\frac{1}{2\omega_{p}}\sin2\varphi_{p}\left(\bknq-\bkq\right)\label{eq:RectCos-Q}
\end{flalign}
\end{ceqn}

\subsection{Deriving Special Cases: \textmd{\normalsize{}$\protect\Sif{\sin2\omega_{p}t}$
and }$\protect\Sqf{\sin2\omega_{p}t}$}

\subsubsection{Integration of $I$-bits}

We derive the result of bit pulse integration for this special case.
\begin{flalign*}
 & \Sif{\sin2\omega_{p}t}\\
 & =\IntI b_{I}\left(t-\tau\right)\sin2\omega_{p}t\: dt\\
 & =\bkni\left[-\frac{1}{2\omega_{p}}\cos2\omega_{p}t\vphantom{\prod}\right]_{\left(2k-1\right)T}^{\left(2k-1\right)T+\underline{\tau}}+\bki\left[-\frac{1}{2\omega_{p}}\cos2\omega_{p}t\vphantom{\prod}\right]_{\left(2k-1\right)T+\underline{\tau}}^{\left(2k+1\right)T}\\
 & =-\frac{\bkni}{2\omega_{p}}\left[\cos2\omega_{p}t\vphantom{\prod}\right]_{\left(2k-1\right)T}^{\left(2k-1\right)T+\underline{\tau}}-\frac{\bki}{2\omega_{p}}\left[\cos2\omega_{p}t\vphantom{\prod}\right]_{\left(2k-1\right)T+\underline{\tau}}^{\left(2k+1\right)T}
\end{flalign*}
Performing the integration results in (we denote $\omega_{p}\tu=\underline{\varphi_{p}}$):
\begin{flalign*}
 & =-\frac{\bkni}{2\omega_{p}}\left[\cos\left(\left(2k-1\right)\pi+2\underline{\varphi_{p}}\right)-\cos\left(\left(2k-1\right)\pi\right)\right]\\
 & \hphantom{=}-\frac{\bki}{2\omega_{p}}\left[\cos\left(\left(2k+1\right)\pi\right)-\cos\left(\left(2k-1\right)\pi+2\underline{\varphi_{p}}\right)\right]\\
 & =-\frac{\bkni}{2\omega_{p}}\left(\cos\left(-\pi+2\underline{\varphi_{p}}\right)-\cos\left(-\pi\right)\right)-\frac{\bki}{2\omega_{p}}\left(\cos\pi-\cos\left(-\pi+2\underline{\varphi_{p}}\right)\right)\\
 & =-\frac{\bkni}{2\omega_{p}}\left(1-\cos2\underline{\varphi_{p}}\right)+\frac{\bki}{2\omega_{p}}\left(1-\cos2\underline{\varphi_{p}}\right)\\
 & =-\frac{1}{2\omega_{p}}\left(1-\cos2\underline{\varphi_{p}}\right)\left(\bkni-\bki\right)\\
 & \intertext{\text{Using \ensuremath{\cos}2\ensuremath{\underline{\varphi_{p}}} = \ensuremath{\cos\left(2\omega_{p}\left(\tau-2k_{\tau}T\right)\right)} = \ensuremath{\cos\left(2\varphi_{p}-2k_{\tau}\pi\right)} = \ensuremath{\cos}2\ensuremath{\varphi_{p}}}}\\
 & =-\frac{1}{2\omega_{p}}\left(1-\cos2\varphi_{p}\right)\left(\bkni-\bki\right)
\end{flalign*}
The overall result is \begin{ceqn} 
\begin{flalign}
\Sif{\sin2\omega_{p}t} & =-\frac{1}{2\omega_{p}}\left(1-\cos2\varphi_{p}\right)\left(\bkni-\bki\right)\label{eq:RectSin-I}
\end{flalign}
\end{ceqn}

\subsubsection{Integration of $Q$-bits}

This case can be performed analogously to \secref{Cos-I-in-Q}, with
the following two simplifications:

\begin{flalign*}
\cos2\underline{\varphi_{p}^{Q}} & =\cos2\omega_{p}\tuq=\cos\left(2\frac{\pi}{2T}\left(\tau^{Q}-2k_{\tau}^{Q}T\right)\right)=\cos\left(\frac{\tau^{Q}\pi}{T}-2k_{\tau}^{Q}\pi\right)=\cos2\varphi_{p}^{Q}\\
 & \intertext{\text{and }}\\
\cos2\varphi_{p}^{Q} & =\cos\left(2\varphi_{p}+\pi\right)=-\cos2\varphi_{p}
\end{flalign*}
The overall result is \begin{ceqn} 
\begin{flalign}
\Sqf{\sin2\omega_{p}\left(t\right)} & =-\frac{1}{2\omega_{p}}\left(1+\cos2\varphi_{p}\right)\left(\bknq-\bkq\right)\label{eq:RectSin-Q}
\end{flalign}
\end{ceqn}

\section{Demodulator Output for Signals with Both Offsets $\tau,\varphi_{c}$}

\label{apx:Offset-Both-I}With the tools presented in \apxref{IntegratingRect},
we can now proceed to prove \thmref{FullyUSync}. \setcounter{thm}{0}
\begin{thm}
For an interfering MSK signal $u\left(t\right)$ with parameters $\tau$
and $\varphi_{c}$, the contribution to the demodulation output $\Lambda_{u}^{I}\left(k\right)$
is given by 
\begin{flalign*}
\Lambda_{u}^{I}\left(k\right) & =\frac{1}{4}A_{u}\left\{ \cos\varphi_{c}\left[\cos\varphi_{p}\left(\tu\bkni+\left(2T-\tu\right)\bki\right)-\frac{2T}{\pi}\sin\varphi_{p}\left(\bkni-\bki\right)\right]\right.\\
 & \hphantom{=\frac{1}{4}A_{u}}\left.-\sin\varphi_{c}\left[\sin\varphi_{p}\left(\tuq\bknq+\left(2T-\tuq\right)\bkq\right)+\frac{2T}{\pi}\cos\varphi_{p}\left(\bknq-\bkq\right)\right]\right\} .
\end{flalign*}
\end{thm}
\begin{IEEEproof}
We first derive the resulting signal after demodulation (\eqref{Lambda}).
\begin{flalign*}
 & u\left(t\right)\phi_{I}\left(t\right)\\
 & =A_{u}\left[b_{I}\left(t-\tau\right)\cos\left(\omega_{p}t-\varphi_{p}\right)\cos\left(\omega_{c}t+\varphi_{c}\right)\right.\\
 & \hphantom{=A_{u}}\left.+b_{Q}\left(t-\tau\right)\sin\left(\omega_{p}t-\varphi_{p}\right)\sin\left(\omega_{c}t+\varphi_{c}\right)\right]\left[\cos\omega_{p}t\cos\omega_{c}t\right]\\
 & =A_{u}\left[\left(b_{I}\left(t-\tau\right)\cos\left(\omega_{p}t-\varphi_{p}\right)\cos\omega_{p}t\cos\left(\omega_{c}t+\varphi_{c}\right)\cos\omega_{c}t\right)\right.\\
 & \hphantom{=A_{u}}\left.+\left(b_{Q}\left(t-\tau\right)\sin\left(\omega_{p}t-\varphi_{p}\right)\cos\omega_{p}t\sin\left(\omega_{c}t+\varphi_{c}\right)\cos\omega_{c}t\right)\right]\\
 & =\frac{A_{u}}{4}\left[\left(b_{I}\left(t-\tau\right)\left(\cos\varphi_{p}+\cos\left(2\omega_{p}t-\varphi_{p}\right)\right)\left(\cos\varphi_{c}+\cos\left(2\omega_{c}t+\varphi_{c}\right)\right)\right)\right.\\
 & \hphantom{=\frac{A_{u}}{4}}\left.+\left(b_{Q}\left(t-\tau\right)\left(\sin\left(-\varphi_{p}\right)+\sin\left(2\omega_{p}t-\varphi_{p}\right)\right)\left(\sin\varphi_{c}+\sin\left(2\omega_{c}t+\varphi_{c}\right)\right)\right)\right]\\
 & \intertext{\text{We apply perfect lowpass filtering (\ensuremath{\star}) to filter out high-frequency components (\ensuremath{2\omega_{c}t})}}\\
 & \overset{\star}{=}\frac{A_{u}}{4}\left[\left(b_{I}\left(t-\tau\right)\cos\varphi_{c}\left(\cos\varphi_{p}+\cos\left(2\omega_{p}t-\varphi_{p}\right)\right)\right)\right.\\
 & \hphantom{=\frac{A_{u}}{4}}\left.+\left(b_{Q}\left(t-\tau\right)\sin\varphi_{c}\left(\sin\left(2\omega_{p}t-\varphi_{p}\right)-\sin\varphi_{p}\right)\right)\right]\\
 & =\frac{A_{u}}{4}\left[\left(b_{I}\left(t-\tau\right)\cos\varphi_{c}\left(\cos\varphi_{p}+\cos2\omega_{p}t\cos\varphi_{p}+\sin2\varphi_{p}t\sin\varphi_{p}\right)\right)\right.\\
 & \hphantom{=\frac{A_{u}}{4}}\left.+\left(b_{Q}\left(t-\tau\right)\sin\varphi_{c}\left(-\sin\varphi_{p}+\sin2\omega_{p}t\cos\varphi_{p}-\cos2\omega_{p}t\sin\varphi_{p}\right)\right)\right]
\end{flalign*}
The bit decision is performed by integration over the bit interval
$k$.
\begin{flalign*}
 & \IntI u\left(t\right)\phi_{I}\left(t\right)dt\\
 & =\frac{A_{u}}{4}\left[\cos\varphi_{c}\IntI b_{I}\left(t-\tau\right)\left(\cos\varphi_{p}+\cos2\omega_{p}t\cos\varphi_{p}+\sin2\varphi_{p}t\sin\varphi_{p}\right)dt\right.\\
 & \hphantom{=\frac{A_{u}}{4}}\left.+\sin\varphi_{c}\IntI b_{Q}\left(t-\tau\right)\left(-\sin\varphi_{p}+\sin2\omega_{p}t\cos\varphi_{p}-\cos2\omega_{p}t\sin\varphi_{p}\right)dt\right]\\
 & =\frac{A_{u}}{4}\left[\cos\varphi_{c}\mathcal{X}_{1}+\sin\varphi_{c}\mathcal{X}_{2}\right]
\end{flalign*}
We derive the results for both terms $\mathcal{X}_{1}$ and $\mathcal{X}_{2}$
individually in the following two sections.

Putting the two results in \eqref{X1} and \eqref{X2} together, the
overall result is \begin{ceqn} 
\begin{flalign*}
 & \IntI u\left(t\right)\phi_{I}\left(t\right)dt\\
 & =\frac{A_{u}}{4}\left\{ \cos\varphi_{c}\left[\cos\varphi_{p}\left(\underline{\tau}\bkni+\left(2T-\underline{\tau}\right)\bki\right)-\frac{2T}{\pi}\sin\varphi_{p}\left(\bkni-\bki\right)\right]\right.\\
 & \hphantom{=\frac{A_{u}}{4}}\left.-\sin\varphi_{c}\left[\sin\varphi_{p}\left(\tuq\bknq+\left(2T-\tuq\right)\bkq\right)+\frac{2T}{\pi}\cos\varphi_{p}\left(\bknq-\bkq\right)\right]\right\} 
\end{flalign*}
\end{ceqn}
\end{IEEEproof}

\subsection{Integrating the Term $\mathcal{X}_{1}$}

\begin{flalign*}
 & \IntI b_{I}\left(t-\tau\right)\left(\cos\varphi_{p}+\cos2\omega_{p}t\cos\varphi_{p}+\sin2\varphi_{p}t\sin\varphi_{p}\right)dt\\
 & =\cos\varphi_{p}\IntI b_{I}\left(t-\tau\right)dt+\cos\varphi_{p}\IntI b_{I}\left(t-\tau\right)\cos2\omega_{p}t\: dt\\
 & \hphantom{=}+\sin\varphi_{p}\IntI b_{I}\left(t-\tau\right)\sin2\omega_{p}t\: dt\\
 & =\cos\varphi_{p}\Sif 1+\cos\varphi_{p}\Sif{\cos2\omega_{p}t}+\sin\varphi_{p}\Sif{\sin2\omega_{p}t}
\end{flalign*}
By using the results in \apxref{IntegratingRect} (\eqref{Rect1-I,RectCos-I,RectSin-I}),
we can reformulate this equation to 
\begin{flalign*}
 & =\cos\varphi_{p}\left(\underline{\tau}\bkni+\left(2T-\underline{\tau}\right)\bki\right)\\
 & \hphantom{=}-\frac{\bkni}{2\omega_{p}}\left(\cos\varphi_{p}\sin2\varphi_{p}+\sin\varphi_{p}\left(1-\cos2\varphi_{p}\right)\right)+\frac{\bki}{2\omega_{p}}\left(\cos\varphi_{p}\sin2\varphi_{p}+\sin\varphi_{p}\left(1-\cos2\varphi_{p}\right)\right)
\end{flalign*}
Simplifying this equation yields the desired result. 
\begin{flalign*}
 & =\cos\varphi_{p}\left(\tu\bkni+\left(2T-\tu\right)\bki\right)-\left(\frac{\bkni-\bki}{2\omega_{p}}\right)\left(\sin2\varphi_{p}\cos\varphi_{p}-\cos2\varphi_{p}\sin\varphi_{p}+\sin\varphi_{p}\right)\\
 & =\cos\varphi_{p}\left(\tu\bkni+\left(2T-\tu\right)\bki\right)-\frac{\sin\varphi_{p}}{\omega_{p}}\left(\bkni-\bki\right)\\
 & =\cos\varphi_{p}\left(\tu\bkni+\left(2T-\tu\right)\bki\right)-\frac{2T}{\pi}\sin\varphi_{p}\left(\bkni-\bki\right)
\end{flalign*}
In the second step in the previous derivation, we used the following
simplification:
\begin{flalign*}
 & \sin2\varphi_{p}\cos\varphi_{p}-\cos2\varphi_{p}\sin\varphi_{p}+\sin\varphi_{p}\\
 & =2\cos^{2}\varphi_{p}\sin\varphi_{p}-\left(2\cos^{2}\varphi_{p}-1\right)\sin\varphi_{p}+\sin\varphi_{p}\\
 & =\left(2\cos^{2}\varphi_{p}-2\cos^{2}\varphi_{p}+1+1\right)\sin\varphi_{p}\\
 & =2\sin\varphi_{p}
\end{flalign*}
Overall, the result is \begin{ceqn}
\begin{align}
\mathcal{X}_{1} & =\cos\varphi_{p}\left(\tu\bkni+\left(2T-\tu\right)\bki\right)-\frac{2T}{\pi}\sin\varphi_{p}\left(\bkni-\bki\right)\label{eq:X1}
\end{align}
\end{ceqn}

\subsection{Integrating the Term $\mathcal{X}_{2}$}

We will now derive the second integral. We must use the rules for
$Q$ pulse integration with $I$ intervals (\apxref{Int-Qbits}).
\begin{flalign*}
 & \IntI b_{Q}\left(t-\tau\right)\left(-\sin\varphi_{p}-\cos2\omega_{p}t\sin\varphi_{p}+\sin2\omega_{p}t\cos\varphi_{p}\right)dt\\
 & =-\left[\IntI b_{Q}\left(t-\tau\right)\sin\varphi_{p}dt+\IntI b_{Q}\left(t-\tau\right)\cos2\omega_{p}t\sin\varphi_{p}dt\right.\\
 & \hphantom{=-}\left.-\IntI b_{Q}\left(t-\tau\right)\sin2\omega_{p}t\cos\varphi_{p}dt\right]\\
 & =-\left[\sin\varphi_{p}\Sqf 1+\sin\varphi_{p}\Sqf{\cos2\omega_{p}t}-\cos\varphi_{p}\Sqf{\sin2\omega_{p}t}\right]
\end{flalign*}
By using the results in \apxref{IntegratingRect} (\eqref{Rect1-Q,RectCos-Q,RectSin-Q}),
we can reformulate this equation to 
\begin{flalign*}
 & =-\sin\varphi_{p}\left(\tuq\bknq+\left(2T-\tuq\right)\bkq\right)\\
 & \hphantom{=}-\frac{1}{2\omega_{p}}\left(\sin\varphi_{p}\sin2\p+\cos\varphi_{p}\left(1+\cos2\p\right)\right)\left(\bknq-\bkq\right)\\
 & \intertext{\text{Simplifying yield the desired result}}\\
 & =-\sin\varphi_{p}\left(\tuq\bknq+\left(2T-\tuq\right)\bkq\right)\\
 & \hphantom{=}-\frac{1}{2\omega_{p}}\left(\sin2\varphi_{p}\sin\varphi_{p}+\cos2\varphi_{p}\cos\varphi_{p}+\cos\varphi_{p}\right)\left(\bknq-\bkq\right)\\
 & =-\left[\sin\varphi_{p}\left(\tuq\bknq+\left(2T-\tuq\right)\bkq\right)+\frac{2T}{\pi}\cos\varphi_{p}\left(\bknq-\bkq\right)\right]
\end{flalign*}
In the last step, we used the following simplification 
\begin{flalign*}
 & \sin2\varphi_{p}\sin\varphi_{p}+\cos\varphi_{p}+\cos2\varphi_{p}\cos\varphi_{p}\\
 & =2\sin^{2}\varphi_{p}\cos\varphi_{p}+\cos\varphi_{p}+\left(1-2\sin^{2}\varphi_{p}\right)\cos\varphi_{p}\\
 & =\left(2\sin^{2}\varphi_{p}+1+1-2\sin^{2}\varphi_{p}\right)\cos\varphi_{p}\\
 & =2\cos\varphi_{p}
\end{flalign*}
Overall, the result is \begin{ceqn} 
\begin{align}
\mathcal{X}_{2} & =-\left[\sin\varphi_{p}\left(\tuq\bknq+\left(2T-\tuq\right)\bkq\right)+\frac{2T}{\pi}\cos\varphi_{p}\left(\bknq-\bkq\right)\right]\label{eq:X2}
\end{align}
\end{ceqn}

\end{document}